\documentclass[draftclsnofoot, onecolumn, 12pt]{IEEEtran}
  \usepackage[dvips]{graphicx}
  \graphicspath{{./}}
  \DeclareGraphicsExtensions{.eps}
  \usepackage{subfigure}
\usepackage{color}
\usepackage{url}

\usepackage{amsmath}
\usepackage{amsthm}
\usepackage{amsfonts}
\usepackage{amssymb}
\usepackage{bm}
\newtheorem{lemma}{Lemma}
\newtheorem{theorem}{Theorem}
\newtheorem{definition}{Definition}

\newtheorem{remark}{Remark}
\newtheorem{prob}{Problem}

\usepackage{algorithmic}
\usepackage{array}
\usepackage{threeparttable}

\usepackage{enumerate}

\begin{document}
%
\title{Closed-Form Delay-Optimal Computation Offloading in Mobile Edge Computing Systems}

\author
{\IEEEauthorblockN{Xianling Meng,~\IEEEmembership{Student Member,~IEEE}, Wei Wang,~\IEEEmembership{Senior Member,~IEEE}, \\
Yitu Wang,~\IEEEmembership{Student Member,~IEEE},
Vincent K. N. Lau,~\IEEEmembership{Fellow,~IEEE},
 Zhaoyang~Zhang,~\IEEEmembership{Member,~IEEE}}

\thanks{This work has been submitted to the IEEE for possible publication. Copyright may be transferred without notice, after which this version may no longer be accessible.}
\thanks{Part of this work has been presented in IEEE Globecom 2018, Abu Dhabi, UAE. \cite{conference}}
\thanks{Xianling Meng, Wei Wang, Yitu Wang and Zhaoyang Zhang are with College of Information Science and Electronic Engineering, Zhejiang University, Hangzhou 310027, China.}
\thanks{Vincent K. N. Lau is with Department of Electrical and Computer Engineering, Hong Kong University of Science and Technology, Hong Kong.}
}


%


\maketitle

\begin{abstract}
	Mobile edge computing (MEC) has recently emerged as a promising technology to release the tension between computation-intensive applications and resource-limited mobile terminals (MTs). In this paper, we study the delay-optimal computation offloading in computation-constrained MEC systems.
	We consider the computation task queue at the MEC server due to its constrained computation capability. In this case, the task queue at the MT and that at the MEC server are strongly coupled in a cascade manner, which creates complex interdependencies and brings new technical challenges.
	We model the computation offloading problem as an infinite horizon average cost Markov decision process (MDP), and approximate it to a virtual continuous time system (VCTS) with reflections. Different to most of the existing works, we develop the dynamic instantaneous rate estimation for deriving the closed-form approximate priority functions in different scenarios. Based on the approximate priority functions, we propose a closed-form multi-level water-filling computation offloading solution to characterize the influence of not only the local queue state information (LQSI) but also the remote queue state information (RQSI). A extension is provided from single MT single MEC server scenarios to multiple MTs multiple MEC servers scenarios and several insights are derived. Finally, the simulation results show that the proposed scheme outperforms the conventional schemes.
\end{abstract}

\IEEEpeerreviewmaketitle

\newpage
\section{Introduction}

Smart mobile terminals (MTs) with advanced communication and computation capabilities facilitate us with a pervasive and powerful platform to realize many emerging computation-intensive mobile applications, e.g., interactive gaming, character recognition, and natural language processing \cite{YM2016}, \cite{BF2016}. These pose exigent requirements on the quality of computation experience, especially for the delay-sensitive applications.

{Computation offloading~\cite{PM2017}, which offloads the computation tasks to the offloading destination, is one of the fundamental services to improve the computation performance, i.e., delay performance. In computation offloading services, both the communication capability of the MT and the computation capability of the offloading destination will influence the delay performance. Specifically,
	\begin{itemize}
		\item \textbf{The communication capability of the MT:} 
		The offloading rate varies according to the time-varying wireless channel quality between the MT and the offloading destination.	Poor communication capabilities will result in the starvation of the computation of the offloading destination, which induces a large queuing delay at the MT.
		\item \textbf{The computation capability of the offloading destination:} 
		In practical scenarios, the offloaded tasks cannot be executed immediately because the computation capability of the offloading destination is not infinity. Both the computation time and the waiting time at the offloading destination will influence the delay performance.
\end{itemize}}

In~\cite{AG2006}, a basic two-party communication complexity model is studied for the networked computation problems, with a particular emphasis on the communication aspect of computation. In~\cite{MM2014}, the communication and computation capabilities are jointly optimized to minimize the delay under the constraint of energy consumption. Since the cloud computing servers are usually computationally powerful, it is reasonable to neglect the executing time at the server. However, the remote cloud computing servers are always far away from the MTs and the large communication delay cannot be decreased, so the cloud computation offloading is not fit for the delay-sensitive applications.

Mobile edge computing (MEC)~\cite{HL2018} is emerged as a promising technology to handle the explosive computation demands and the everincreasing computation quality requirements. Different from conventional cloud computing systems, MEC offers computation capability in close proximity to the MT. Therefore, by offloading the computation task from the MTs to the MEC servers, the delay performance can be greatly improved \cite{WS2016} \cite{TX2017} \cite{PC2016}.

{Most aforementioned works focus on optimize the local computation delay or the communication delay and neglect the computation delay at the MEC server. However, for computation-constrained MEC system, the computation capability of the MEC server is limited in MEC systems and neglecting the computation delay at the MEC server will lead to the deviation from the optimality. A computation offloading policy is strongly desired for computation-constrained MEC systems to achieve superior delay performance.}

{In this paper, we aim to achieve a delay-optimal computation offloading policy for computation-constrained MEC systems. Specially, the computation offloading policy will consider not only the current computation task delay, but also the future delay performance of the MEC system for superior delay performance.
	To investigate the optimal delay performance, a systematic approach to the delay-aware optimization problem is through the Markov decision process (MDP), but there are a couple of technical challenges involved as follows:
	\begin{itemize}
		\item \textbf{Challenges due to the Cascade Queue Coupling:}
		Because of the cascade manner between the local task queue and the remote task queue, the offloading policy should be adapted to not only the channel state information (CSI) and the local queue state information (LQSI), but also the remote queue state information (RQSI) with the practical consideration on the limited computation capability of the MEC server.
		Specifically, for achieving the delay-optimal computation offloading, we need jointly consider the queue lengths of the LQSI and the RQSI, and choose the efficient transmission opportunities to execute the offloading based on the time-varying CSI.
		Also, for fully using the computation capability of both the MT and the MEC server, we need to maintain the balance between two cascade queues by adjusting the transmission rate (power), because the departure of the local task queue is the arrival of the remote task queue.
		\item \textbf{Challenges due to the closed-form MDP solution:}
		For obtaining the optimal solution of the MDP optimization problem, a Bellman equation needs to be solved, which is well known as NP-hard, and nontrivial to obtain an optimal solution in closed-form with low computational complexity.
		Also, for maintaining the cascade queue balance, the time-varying system, which consists of the random task arrivals, the local computation, the transmission and the remote task computation, cannot adopt a simple long-term average state formulation.
		The optimal computation offloading policy should have the ability to adapt to the random task arrivals and make sure the cascade queue system can converge to the delay-optimal steady state by adjusting the local computation rate (power) and the transmission rate (power).
		The system dynamics increase the difficulties of solving the formulated Bellman equation.
\end{itemize}}

{For overcoming the aforementioned challenges, we develop an analytical framework for delay-optimal computation offloading in computation-constrained MEC systems, and derive a closed-form offloading policy. Our key contributions are summarized as follows:
	\begin{itemize}
		\item We consider the computation-constrained MEC server for the delay-optimal computation offloading problem. In this system, the computation delay of the MEC server cannot be neglected, and the cascade queue balance should be maintained. For achieving good delay performance, the delay-optimal computation offloading policy should jointly consider the CSI, the LQSI and the RQSI simultaneously.
		\item We formulate the delay-optimal computation offloading problem as an infinite horizon average cost MDP, and adopt a virtual continuous time system (VCTS) with reflections to overcome the curse of dimensionality.	
		Next, we develop a multi-level water-filling computation offloading policy for jointly considering the CSI, the LQSI and the RQSI.
		Then, we derive the dynamic instantaneous rate estimation for maintaining the cascade queue balance by estimating the in-out rate difference of the queue system.
		Finally, we obtain approximate priority functions in both the computation sufficient scenario and the computation constrained scenario.
		\item We extend our policy to the multi-MT multi-server scenario by adopting learning approach. Specifically, we compare the main differences between two scenarios, and derive a computation offloading policy by learning the access ratios from the historical access records.
\end{itemize}}

{The rest of this paper is organized as follows.
	Section II discusses the related works.
	Section III presents the system model and formulates the computation offloading problem.
	Section IV provides the optimality conditions via establishing the VCTS.
	Section V proposes the delay-optimal computation offloading policy {and the dynamic instantaneous rate estimation}.
	Section VI extends the computation offloading policy to the multi-MT multi-server scenarios and derives some brief insights.
	The performance of the proposed policy is evaluated by simulation in Section VII.
	Finally, this paper is concluded in Section VIII.}

\section{Related Works}
Since this paper studies the delay-optimal computation offloading in MEC systems, in this section, we briefly review the existing works on computation offloading and delay-aware considerations.

\subsection{Computation Offloading in MEC Systems}
Computation offloading in MEC systems has attracted significant attentions recently.
In~\cite{WYTOFF21}, the computation tasks are chosen to offload for minimizing the average power consumption. In~\cite{WYTOFF22}, the energy-delay tradeoff is analyzed for single-user MEC systems. Then, the results are extended to multi-user systems in~\cite{WYTOFF3}. In~\cite{WYTOFF23}, a distributed computational offloading algorithm is proposed using game theory. In \cite{WYTOFF24}, both the radio and computational resources are optimized for computation offloading in multi-cell MEC systems.

For delay-sensitive applications, it is necessary to consider the delay performance for computation offloading \cite{YZ2012}. Significant theoretical and experimental research has been done in various areas to show that computation offloading can significantly enhance the delay performance.
In~\cite{EG1-JL2016}, an one-dimensional search algorithm is proposed to minimize the total delay. {In \cite{YM2016}, an offloading strategy based on Lyapunov optimization is adopted to minimize the total cost which consists of delay and energy consumption.}
In \cite{EG3-WL2015}, two offline strategies based on the constrained MDP are proposed to minimize the energy consumption under a delay constraint. 
In \cite{EG4-XC2016}, a distributed computation offloading algorithm is proposed to achieve Nash equilibrium between delay and energy consumption. In \cite{EG9-OM2015}, joint communication-computation optimization are studied to minimize the delay and energy consumption.

However, the above existing works take the assumption that the MEC server is computationally powerful enough such that the offloaded computation tasks are executed immediately once arriving the server. 
In this paper, we consider the limited computation capability of the MEC server, and include the queuing time at the MEC server into the delay performance of computation. In this case, we handle the coupling between the computation capability of the MEC server and the communication capability of the MT, and propose a computation offloading policy to balance the communication-computation tradeoff.

\subsection{Delay-Aware Considerations}
To optimize the delay performance, there are several common approaches to handle delay-aware
resource allocation \cite{WYTOFF15}. Large deviation \cite{WYTOFF16} is an approach to convert the delay constraint into an equivalent rate constraint. However, this method achieves good delay performance only in a large delay regime. Stochastic majorization \cite{WYTOFF17} provides a way to minimize the delay for the cases with symmetric arrivals. Lyapunov optimization \cite{WYTOFF11} is an effective approach on queue stability, 
{but it is effective only when the queue backlog is large.}

MDP \cite{DP2005} is a systematic approach to minimize the delay.
In general, the optimal control policy can be obtained by solving the well-known Bellman equation.
Conventional solutions to the Bellman equation, such as brute-force value iteration or policy iteration \cite{DP2005}, have huge complexity (i.e., the curse of dimensionality), because solving the Bellman equation involves solving an exponentially large system of non-linear equations. There are some existing works that use the stochastic approximation approach with distributed online learning algorithm~\cite{YC2011}, which has linear complexity. However, the stochastic learning approach can only give a numerical solution to the Bellman equation and may suffer from slow convergence and lack of insight~\cite{WW2015}.

In this paper, we address this issue head-on by transforming the discrete time MDP to a continuous time VCTS with reflections, such that it is possible to derive a closed-form computation offloading policy by solving the stochastic differential equations.

\begin{figure*}
	\centering
	\includegraphics[width=0.9\textwidth]{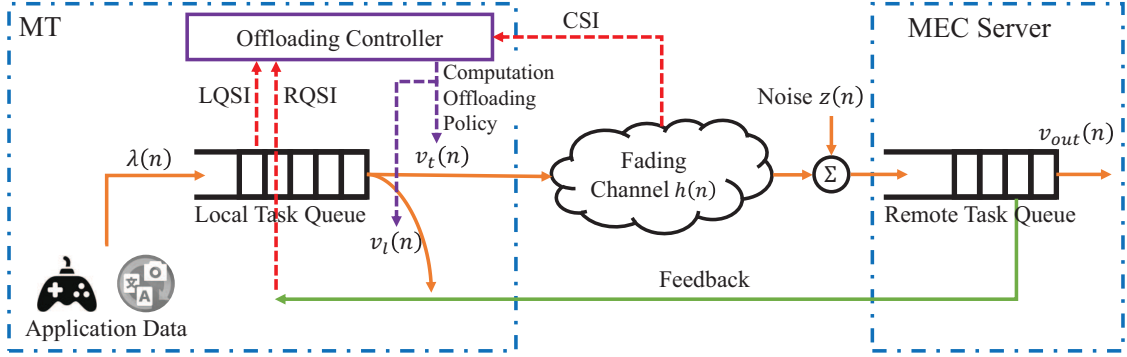}
	\caption{Cascade queue system}
	\label{fig:systemmodel}
\end{figure*}

\section{System Model and Problem Formulation}
In this section, we introduce a MEC system with bursty task arrivals.
First, we elaborate the system model and introduce the queue dynamics at both the MT and at the MEC server. Then, we define the computation offloading policy and formulate the delay-optimal optimization problem. 

\subsection{MEC System Model}
Consider a MEC system with one MT and one MEC server, as shown in Fig. \ref{fig:systemmodel}.
The MT executes its computation tasks with two approaches, including the local computation at the MT and the computation offloading from the MT to the MEC server.
In our system, time is slotted with duration $\tau$, and the slots are indexed by $n$.

First, we consider the approach that the computation tasks are computed at the MT.
With dynamic voltage and frequency scaling (DVFS) techniques, the local computation rate can be adjusted by changing the CPU-cycle frequencies \cite{WZ2013}. Denote $f_{MT}$ as the CPU-cycle frequency of the MT, the local computation rate at the $n$-th time slot can be expressed as
\begin{equation}
v_l(n) = k(n)f_{MT},
\label{vl1}
\end{equation}
where $k(n)$ is the scale factor\footnote{By this scale factor, we unify the transmission rate and the computation rate of the MT.} between the packet size and the amount of floating point operations of the computation task with mean $\mathbb{E}[k] = \overline{k}$.

High CPU-cycle frequency increases the power consumption. 
The power consumption for the local computation at the MT is
\begin{equation}
P_l(n) = cf_{MT}^2,
\label{vl2}
\end{equation}
where $c$ is the effective switched capacitance that depends on the CPU architecture. {Based} on (\ref{vl1}) and (\ref{vl2}), $v_l(P_l(n))$ is calculated as
\begin{equation}
v_l(P_l(n)) = \frac{k(n)}{\sqrt{c}}\sqrt{P_l(n)}.
\end{equation}

{Next, we consider the approach that the computation tasks are offloaded to the MEC server. This approach contains the transmission phase at the MT and the computation phase at the MEC server.}

Denote $H(n)$ as the CSI which is the instantaneous channel path gain from the MT to the MEC server at the $n$-th time slot, with mean $\mathbb{E}[H(n)] = L$. Denote $N_0$ as the noise power of the complex Gaussian additive channel and $B$ as the bandwidth. For given CSI $H(n)$ and transmission power $P_t(n)$, the transmission rate of the MT is calculated as
\begin{equation}
v_t(P_t(n), H(n)) = B\log_2\bigg(1+\frac{P_t(n)H(n)}{N_0}\bigg).
\end{equation}

Denote $f_{MEC}$ as the CPU-cycle frequency of the MEC server. The computation rate of the MEC server at the $n$-th time slot can be expressed as 
\begin{equation}
v_{out}(n) = k(n)f_{MEC},
\label{vout}
\end{equation}
with mean $\mathbb{E}[v_{out}] = \overline{v_{out}}$.

\subsection{Queue Dynamics}

To analyze the delay performance, we first discuss the local and remote task queues.
Let $Q_l(n), Q_r(n) \in [0, \infty)$ denote the LQSI (packets) at the MT and the RQSI (packets) at the MEC server at the beginning of the $n$-th time slot, respectively.
Let $\lambda (n)\tau$ be the random arrivals of computation tasks (packets) at the end of the $n$-th time slot at the MT. Assume that $\lambda (n)$ is i.i.d over time slots, with $\mathbb{E}[\lambda]=\overline{\lambda}$, where $\overline{\lambda}$ is the average task arrival rate.
Hence, the dynamics of the local task queue at the MT is given by
\begin{equation}
Q_l(n+1) = [Q_l(n) - v_t(P_t(n), H(n))\tau - v_l(P_l(n))\tau]^+ + \lambda (n)\tau,
\label{eq:localqueuedis}
\end{equation}
and that of the remote task queue at the MEC server is given by
\begin{equation}
Q_r(n+1) = [Q_r(n) - v_{out}(n)\tau]^+ + v_t(P_t(n), H(n))\tau,
\label{eq:remotequeuedis}
\end{equation}
where $[\cdot]^+ \triangleq \max\{\cdot, 0\}$.

Fig. \ref{fig:systemmodel} illustrates the queue system for computation offloading, where the CSI, the LQSI and the RQSI are jointly considered to make an appropriate computation offloading decision.

\begin{remark}[Cascade Coupling of Local and Remote Queues]
	The local queue dynamics in (\ref{eq:localqueuedis}) and the remote queue dynamics in (\ref{eq:remotequeuedis}) are coupled together by a cascade control, because the departure of the former is the arrival of the latter. This cascade coupling creates complex interdependence and makes the computation offloading problem an involved stochastic optimization problem.\hfill\IEEEQED
\end{remark}

\subsection{Computation Offloading Policy}
{Next, we define the computation offloading policy for the mentioned MEC system.} For notation convenience, denote $S(n) = (Q_l(n), Q_r(n), H(n))$ as the state set. The action set consists of $P_l(n)$ and $P_t(n)$.
{At the beginning of the $n$-th time slot, the MT determines the computation offloading action based on the following policy.}
\begin{definition}[Computation Offloading Policy]
	{A computation offloading policy $\Omega$ specifies the offloading actions $P_l(n)$ and $P_t(n)$ that the MT will choose when in state $S(n)$, which the actions are adaptive to all the information $S(i)$ up to time $n$ (i.e., $\{S(i): 0 \leq i \leq n\}$).\hfill\IEEEQED}
	\label{def:pcp}
\end{definition}

Given an offloading policy $\Omega$, the random process $\{S(n)\}$ is a controlled Markov chain with the following transition probability:
\begin{equation}
	\begin{aligned}
	\mathrm{Pr}\big[S(n+1)\mid S(n), \Omega(S(n))\big]
	&=\mathrm{Pr}\big[H(n+1)\big]\cdot \mathrm{Pr}\big[Q_l(n+1) \mid Q_l(n), H(n), \Omega(S(n))\big] \\
	&\qquad\cdot \mathrm{Pr}\big[Q_r(n+1) \mid Q_l(n), Q_r(n), H(n), \Omega(S(n))\big],
	\end{aligned}
\end{equation}
{where the transition probability of the CSI is independent. The probability of the LQSI is based on the last state of the LQSI and the CSI. The probability of the RQSI is related to not only the last state of the RQSI and the CSI, but also the last state of the LQSI because the actual transmission amount cannot exceed $Q_l(n)$. Specifically, The probability $\mathrm{Pr}\big[Q_l(n+1) \mid Q_l(n), H(n), \Omega(S(n))\big]$ is given by}
\begin{eqnarray}
	&&\mathrm{Pr}\big[Q_l(n+1) \mid Q_l(n), H(n), \Omega(S(n)) = \{P_l(n), P_t(n)\}\big]\nonumber\\
	&&=\left \{
	\begin{aligned}
	&\mathrm{Pr}(\lambda (n)), &&\mathrm{if}\; Q_l(n+1)=[Q_l(n) - v_l(P_l(n))\tau - v_t(P_t(n), H(n))\tau]^++\lambda(n)\tau,\\
	&0, &&\mathrm{otherwise}.\nonumber\\
	\end{aligned}
	\right.
\end{eqnarray}
Similarly,  {the probability $\mathrm{Pr}\big[Q_r(n+1) \mid Q_l(n), Q_r(n), H(n), \Omega(S(n))\big]$ is given by}
\begin{eqnarray}
	&&\mathrm{Pr}\big[Q_r(n+1)\mid Q_l(n), Q_r(n), H(n), \Omega(S(n)) = \{P_l(n), P_t(n)\}\big]\nonumber\\
	&&=\left \{
	\begin{aligned}
	&\mathrm{Pr}(v_{out}(n)), &&\mathrm{if}\; Q_l(n)-v_l(P_l(n))\tau - v_t(P_t(n), H(n))\tau \geq 0,\\ &&&\;\;\;\mathrm{and}\;  Q_r(n+1)=[Q_r(n) - v_{out}(n)\tau]^++v_t(P_t(n), H(n))\tau,\\
	&0, &&\mathrm{otherwise}.\nonumber\\
	\end{aligned}
	\right.
\end{eqnarray}
Furthermore, we have the following definition of the admissible offloading policy, which guarantees that the system will converge to a unique steady state.  
\begin{definition}[Admissible Offloading Policy]
	A policy $\Omega$ is admissible if the following requirements are satisfied:
	\begin{itemize}
		\item $\Omega$ is a unichain policy, i.e., the controlled Markov chain $\{S(n)\}$ under $\Omega$ has a single recurrent class (and possibly some transient states).
		\item The queues in the MEC system under $\Omega$ are steady in the sense that $\lim_{n \to \infty}\mathbb{E}^{\Omega}[Q_l^2(n)]<\infty$ and $\lim_{t \to \infty}\mathbb{E}^{\Omega}[Q_r^2(n)]<\infty$, where $\mathbb{E}^{\Omega}$ means taking expectation w.r.t. the probability measure induced by the offloading policy $\Omega$.\hfill\IEEEQED
	\end{itemize}
	\label{def:acp}
\end{definition}

\subsection{Problem Formulation}

Under an admissible offloading policy $\Omega$, the average delay $\overline{D(\Omega)}$ and average power cost $\overline{P(\Omega)}$ starting from a given initial state $S(0)$ are given by
\begin{equation}
\overline{D(\Omega)}=\limsup_{N\to +\infty}\frac{1}{N}\sum_{n=0}^{N-1}\mathbb{E}^{\Omega}[D(n)],
\end{equation}
where {$D(n)$ is denoted as the average queuing delay in slot $n$. 
	For the cascade queue system, the queuing delay of both the local task queue and the remote task queue should be considered.
	Considering the task proportions of local computation and computation offloading are $r_1$ and $r_2$ with $r_1 + r_2 = 1$, we can obtain that the arrival rate of the local task queue and that of the remote task queue are $\lambda$ and $r_2\lambda$, respectively.
	Then the queuing delay $D(n)$ can be expressed\footnote{We aim to develop a delay-optimal computation offloading policy to promote the network performance by adopting the ``packet-level'' delay in this work.} as $D(n) = r_1 \frac{Q_l(n)}{\overline{\lambda}} + r_2 \big[\frac{Q_l(n)}{\overline{\lambda}} + \frac{Q_r(n)}{r_2 \overline{\lambda}}\big] = \frac{1}{\overline{\lambda}}\big[Q_l(n)+Q_r(n)\big]$}, and
\begin{equation}
\overline{P(\Omega)}=\limsup_{N\to +\infty}\frac{1}{N}\sum_{n=0}^{N-1}\mathbb{E}^{\Omega}[P(n)],
\end{equation}
where $P(n) = P_l(n) + P_t(n)$, respectively.

Based on the expressions above, we define the {average cost} $\theta (\Omega)$ for the delay optimization under given weights $\alpha$ and $\beta$ as
\begin{eqnarray}
&\theta (\Omega) &= \alpha \overline{D(\Omega)} + \beta \overline{P(\Omega)}\nonumber\\
&&=\limsup_{N\to +\infty}\frac{1}{N}\sum_{n=0}^{N-1}\mathbb{E}^{\Omega}[g(n)],
\end{eqnarray}
where $g(n) = \alpha D(n) + \beta P(n) = \frac{\alpha}{\overline{\lambda}}\big[Q_l(n)+Q_r(n)\big] + \beta\big[P_l(n) + P_t(n)\big]$.

{Based on the above cost function, we can adjust the weights to satisfy different requirements on average delay or average power. We can achieve the delay-optimal computation offloading policy by solving the following problem:}
\begin{prob}[Delay-Optimal Computation Offloading Problem]
	\begin{equation}
	\min\limits_{\Omega} \theta(\Omega),
	\end{equation}
	where $\Omega$ should satisfy the conditions in  Definition \ref{def:pcp} and Definition \ref{def:acp}.\hfill\IEEEQED
	\label{prob:dopco}
\end{prob}

\section{Optimality Conditions via Virtual Continuous Time System}
In this section, {we first discuss the sufficient optimality condition for Problem \ref{prob:dopco}.} 
As we discussed before, one of the major technical challenges is induced by the huge complexity of solving the multi-dimensional MDP.
{To overcome this challenge, we approximate the problem to a virtual continuous time system (VCTS) with reflections. Based on that, we derive a two-dimensional partial differential equation (PDE) to characterize the priority function.}

\subsection{Optimality Conditions for Problem 1}
Exploiting the i.i.d. property of the CSI, we derive an \textit{equivalent optimality condition} {of Problem~\ref{prob:dopco}} according to \emph{Proposition 4.6.1} in \cite{DP2005} as follows:
\begin{theorem}[Optimality Condition]
	For any given weights $\alpha$ and $\beta$, assume there exists a $(\theta ^*, \{V^*(Q_l, Q_r)\})$ that solves the following equation:
	\begin{equation}
	\theta ^* + V^*(Q_l, Q_r) = \mathbb{E}\bigg[\min\limits_{P_l, P_t}\bigg[g(n) + \sum_{Q'_l, Q'_r}\mathrm{Pr}\big[Q'_l, Q'_r\mid Q_l, Q_r, \Omega(S)\big]V^*(Q'_l, Q'_r)\bigg]\bigg|Q_l, Q_r\bigg].
	\label{eq:bellman}
	\end{equation}
	Furthermore, for all admissible offloading policy $\Omega$ and initial queue state $(Q_l(0), Q_r(0))$, $V^*$ satisfies the following \emph{transversality condition}:
	\begin{equation}
	\lim\limits_{T\to \infty}\frac{1}{T}\mathbb{E}^{\Omega}\big[V^*(Q_l(T), Q_r(T))\mid Q_l(0), Q_r(0)\big] = 0.
	\label{eq:trnasver}
	\end{equation}
	We have the following results:
	\begin{itemize}
		\item $\theta ^ * = \min\limits_{\Omega}\theta (\Omega)$ is the optimal average cost for any initial state $S(0)$ and $V^*(Q_l, Q_r)$ is the priority function.
		\item Suppose there exists an admissible stationary offloading policy $\Omega ^*$ with $\Omega ^*(S) = \{P^*_l, P^*_t\}$ for any $S$, where $\{P^*_l, P^*_t\}$ attains the minimum of the R.H.S. of (\ref{eq:bellman}) for given $S$. Then, the optimal offloading policy of the optimization problem is given by $\Omega ^*$.
	\end{itemize}
	\label{th:SCO}
\end{theorem}
\begin{IEEEproof}
	Please refer to Appendix A.
\end{IEEEproof}

{The solution $V^*(Q_l, Q_r)$ captures the dynamic priority of the task queues for different $(Q_l, Q_r)$. However, obtaining the priority function $V^*(Q_l, Q_r)$ is highly non-trivial because achieving the optimality of the multi-dimensional MDP needs to solve nonlinear fixed point equations. For deriving the closed-form expression, we construct a VCTS with reflections in the following subsection.}

\subsection{Virtual Continuous Time System}
We first define the VCTS, which is a fictitious system with continuous virtual queue state $(q_l(t), q_r(t))$, where $q_l(t), q_r(t) \in [0, \infty)$ are the virtual local queue length and the remote queue length at time $t$ ($t \in [0, \infty)$).

Let $\Omega ^v$ be the virtual computation offloading policy in the VCTS.
Similarly, the virtual offloading policy should be admissible with satisfying the conditions in Definition~\ref{def:acp}.

Given an initial virtual system state $(q_l(0), q_r(0))$ and a virtual policy $\Omega ^v$, the trajectory of the virtual queue system is described by the following coupled differential equations with reflections:
\begin{equation}
\begin{aligned}
&dq_l(t)=\overline{\lambda} dt - \mathbb{E}\big[v_t(P_t(t), H(t))| q_l(t), q_r(t)\big] dt - v_l\big(P_l(t)\big)dt + dRe_l(t) + dRe_t(t),\\
&dq_r(t)=\mathbb{E}\big[v_t(P_t(t), H(t))| q_l(t), q_r(t)\big] dt - \overline{v_{out}} dt - dRe_t(t) + dRe_r(t),
\end{aligned}
\label{eq:dq}
\end{equation}
where $Re_l(t)$ is the reflection process induced by the local computation and associated with the lower boundary $q_l(t) = 0$ for the local task queue, and $Re_t(t)$ is the reflection process induced by the transmission and associated with the lower boundary $q_l(t) = 0$, which are determined by
\begin{equation}
\begin{aligned}
Re_l(t) = \max \bigg\{0, -\min\limits_{t'\leq t}\bigg[ q_l(0) + \int_{0}^{t'}\big(\overline{\lambda} - \mathbb{E}[v_t(P_t(s), &H(s))| q_l(s), q_r(s)]\\
&- v_l(P_l(s))\big)ds \bigg] - \int_{0}^{t'}dRe_t(s)\bigg\}.\\
\end{aligned}
\end{equation}
\begin{equation}
\begin{aligned}
Re_t(t) = \max \bigg\{0, -\min\limits_{t'\leq t}\bigg[ q_l(0) + \int_{0}^{t'}\big(\overline{\lambda} - \mathbb{E}[v_t(P_t(s), &H(s))| q_l(s), q_r(s)]\\
&- v_l(P_l(s))\big)ds \bigg] - \int_{0}^{t'}dRe_l(s)\bigg\}.\\
\end{aligned}
\end{equation}
{$Re_r(t)$ is the reflection process induced by the computation at the MEC server and associated with the lower boundary $q_r(t) = 0$ for the remote task queue, i.e.,}
\begin{equation}
Re_r(t) = \max \bigg\{0, -\min\limits_{t'\leq t} \bigg[q_r(0) - \int_{0}^{t'}dRe_t(s) + \int_{0}^{t'}\big(\mathbb{E}[v_t(P(s), H(s))| q_l(s), q_r(s)] - \overline{v_{out}}\big)ds\bigg]\bigg\},
\end{equation}
where the reflection processes satisfy $Re_l(0) = Re_t(0) = Re_r(0) = 0$.



\subsection{Average Cost Problem Under the VCTS}
For a given admissible virtual offloading policy $\Omega ^v$, we define the average cost of the VCTS from a given initial virtual queue state $(q_l(0), q_r(0))$ as
\begin{equation}
\theta(q_l(0), q_r(0); \Omega ^v) = \lim_{T \to \infty}\frac{1}{T}\int_{0}^{T} \bigg[\frac{\alpha}{\overline{\lambda}} \big(q_l(t) + q_r(t)\big) + \beta \big(P_l(t) + P_t(t)\big)\bigg]dt,
\label{eq:cont_v}
\end{equation}
{then Problem~\ref{prob:dopco} can be reformulated as the following infinite horizon average cost problem in the VCTS:}
\begin{prob}[Infinite Horizon Average Cost Problem in the VCTS]
	\begin{equation}
	\min\limits_{\Omega ^v}\theta(q_l(0), q_r(0); \Omega ^v)
	\end{equation}
	for any given $(q_l(0), q_r(0))$, where $\theta(q_l(0), q_r(0); \Omega ^v)$ is given in (\ref{eq:cont_v}).\hfill\IEEEQED
	\label{prob:vcts}
\end{prob}

This average cost problem has been well-studied in the continuous time optimal control theory~\cite{WYTOFF16}. The solution can be obtained by solving the following \textit{Hamilton-Jacobi-Bellman} (HJB) equation.

\begin{theorem}[Sufficient Optimality Conditions under VCTS]
	Assume there exists a $C^{\infty}$ and a function $V(q_l, q_r)$ of class $\mathcal{C}^1(\mathbb{R}_+^2)$ that satisfy the following HJB equation:
	\begin{equation}
	\begin{aligned}
	\min\limits_{P_l, P_t} \mathbb{E}\bigg[\frac{\alpha}{\overline{\lambda}}\big(q_l + q_r\big) + \beta\big(P_l + P_t\big) + &\frac{\partial V(q_l, q_r)}{\partial q_l}\big[\overline{\lambda} - v_l(P_l) - v_t(P_t, H)\big] \\
	&+ \frac{\partial V(q_l, q_r)}{\partial q_r}\big[v_t(P_t, H) - \overline{v_{out}}\big]\bigg] - C^{\infty} = 0.
	\end{aligned}
	\label{eq:orihjb}
	\end{equation}
	Furthermore, for all admissible virtual control policy $\Omega ^v$ and initial virtual queue state $V(q_l(0), q_r(0))$, the following boundary conditions should be satisfied:
	\begin{eqnarray}
	\left\{
	\begin{aligned}
	&\limsup_{T\to \infty}\int_{0}^{T}\frac{\partial V(0, q_r(t))}{\partial q_l}dRe_l(t) = 0,\\
	&\limsup_{T\to \infty}\int_{0}^{T}\bigg[\frac{\partial V(0, q_r(t))}{\partial q_l} - \frac{\partial V(0, q_r(t))}{\partial q_r}\bigg]dRe_t(t) = 0,\\
	&\limsup_{T\to \infty}\int_{0}^{T}\frac{\partial V(q_l(t), 0)}{\partial q_r}dRe_r(t) = 0,\\
	&\limsup_{T\to \infty}V(q_l(T), q_r(T)) = 0.
	\end{aligned}
	\right.
	\label{eq:bound_cond}
	\end{eqnarray}
	Then we have the following results:
	\begin{itemize}
		\item $C^{\infty} = \min\limits_{\Omega ^v}\theta(q_l(0), q_r(0); \Omega ^v)$ is the optimal average cost, and $V(q_l, q_r)$ is called the \textbf{virtual priority function}.
		\item Suppose there exists an admissible virtual stationary offloading policy $\Omega ^{v*}$ with $\Omega ^{v*}(H, q_l, q_r) = \{P^*_l, P^*_t\}$ for any $(H, q_l, q_r)$, where $\{P^*_l, P^*_t\}$ attains the minimum of the L.H.S. of (\ref{eq:orihjb}) for given $(H, q_l, q_r)$. Then, the optimal offloading policy of Problem \ref{prob:vcts} is given by $\Omega ^{v*}$.
	\end{itemize}
	\label{th:HJB}
\end{theorem}
\begin{IEEEproof}
	Please refer to Appendix B.
\end{IEEEproof}

Similar to \cite{FZ2014}, $V(q_l,q_r)$ in Theorem \ref{th:HJB} can serve as an approximate priority function to the optimal priority function $V^*(Q_l, Q_r)$ in Theorem \ref{th:SCO} with approximation error $o(\tau)$. As a result, solving the Bellman equation (\ref{eq:bellman}) is transformed into a calculus problem of solving the two-dimensional PDE (\ref{eq:orihjb}).

\section{Delay-Optimal Computation Offloading Policy}
{In this section, we solve the two-dimensional HJB equation in Theorem~\ref{th:HJB}.} {By the steady state analyze and the dynamic instantaneous rate estimation of the virtual local and remote queues, we obtain the closed-form solutions to the two-dimensional PDE and extract the insights in different scenarios, including the computation sufficient scenario and the computation constrained scenario. Dynamic instantaneous rate estimation is an important technical approach to achieve the optimal single point and solve the cascade manner MDP framework.}
For simplicity of expression, we denote $V_l = \frac{\partial V(q_l, q_r)}{\partial q_l}$ and $V_r = \frac{\partial V(q_l, q_r)}{\partial q_r}$ in the remaining parts of this paper.

\subsection{Optimal Computation Offloading Structure}
Taking the derivative w.r.t. $P_l$ and $P_t$ on the L.H.S of the HJB equation in (\ref{eq:orihjb}), we obtain the optimal computation offloading in the following theorem:
\begin{theorem}[Optimal Computation Offloading]
	For a given virtual priority function $V(q_l, q_r)$, the optimal computation offloading actions by solving the HJB equation in Theorem \ref{th:HJB} is given by
	\begin{equation}
	P^*_l=\frac{\overline{k}^2}{4c\beta^2}V^2_l,
	\label{eq:plstar}
	\end{equation}
	\begin{equation}
	P^*_t=\bigg(\frac{B}{\beta}(V_l-V_r) - \frac{N_0}{H}\bigg)^+.
	\label{eq:ptstar}
	\end{equation}
	\hfill\IEEEQED
	\label{coro:ptstar}
\end{theorem}

\begin{remark}[Structure of the Optimal Computation Offloading Policy]
	The optimal computation offloading policy in (\ref{eq:ptstar}) depends on the instantaneous CSI, LQSI and RQSI. Furthermore, the optimal offloading transmit power has a \emph{multi-level water-filling} structure, where the water level is adaptive to the LQSI and the RQSI indirectly via the priority function $V(q_l, q_r)$.
	\hfill\IEEEQED
\end{remark}

We then establish the following theorem to substitute the optimal computation offloading policy into the PDE in Theorem \ref{th:HJB} and discuss the sufficient conditions for the existence of solution to the PDE.
\begin{theorem}[PDE with Optimal Computation Offloading Policy]
	With the optimal computation offloading policy in {Theorem~\ref{coro:ptstar}}, the PDE in (\ref{eq:orihjb}) is equivalent to the following PDE:
	\begin{equation}
	\begin{aligned}
	\frac{\alpha}{\overline{\lambda}}\big(q_l + q_r\big) + \beta\big[\frac{\overline{k}^2}{4c\beta^2}V^2_l + \mathrm{EP}(x)\big] - C^{\infty} + V_l\big[\overline{\lambda} - \frac{\overline{k}^2}{2c\beta}V_l - \mathrm{EVP}(x)\big] + V_r\big[\mathrm{EVP}(x) - \overline{v_{out}}\big] = 0,
	\end{aligned}
	\label{eq:simhjb}
	\end{equation}
	where $x=V_l-V_r\geq 0$, $\mathrm{EP}(x) = \frac{Bx}{\beta}\exp\big(-\frac{\beta N_0}{xBL}\big) - \frac{N_0}{L}E_1\big(\frac{\beta N_0}{xBL}\big)$, and $\mathrm{EVP}(x) = BE_1\big(\frac{\beta N_0}{xBL}\big)$.
	After that, there exists a $V(q_l, q_r)$ that satisfies (\ref{eq:bound_cond}) and (\ref{eq:simhjb}) if and only if $\overline{\lambda} = \frac{\overline{k}^2}{2c\beta}V_l + \mathrm{EVP}(x)$, and $\mathrm{EVP}(x) \leq \overline{v_{out}}$.
	\label{th:simhjb}
\end{theorem}
\begin{IEEEproof}
	Please refer to Appendix C.
\end{IEEEproof}

From now on, the main challenge is to find a priority function $V(q_l, q_r)$ that satisfies the PDE in (\ref{eq:simhjb}) and the corresponding boundary conditions in (\ref{eq:bound_cond}).

\subsection{Asymptotic Closed-Form Priority Function}
The PDE in (\ref{eq:simhjb}) is a two-dimensional PDE, which has no closed-form solution for the priority function $V(q_l, q_r)$. In this subsection, we consider the asymptotic analysis under the sufficient conditions for obtaining the closed-form solution of $V(q_l, q_r)$. 

We first analyze the steady states in different cases in the following theorem:
\begin{theorem}[Steady State Analysis] 
	Let $x_e$ be the equilibrium point that $\mathrm{EVP}(x_e) = \overline{v_{out}}$. There exists two possible steady states as follows:
	\begin{enumerate}
		\item 	If $\overline{\lambda} < \mathrm{EVP}(x_e)+\frac{\overline{k}^2}{2c\beta}x_e$, the steady state should satisfy $\overline{\lambda} = \frac{\overline{k}^2}{2c\beta}V_l + \mathrm{EVP}(x)$ and  $\mathrm{EVP}(x) < \overline{v_{out}}$ with $x = V_l$.
		
		\item If $\overline{\lambda} \geq \mathrm{EVP}(x_e)+\frac{\overline{k}^2}{2c\beta}x_e$, the steady state should satisfy $\overline{\lambda} = \frac{\overline{k}^2}{2c\beta}V_l + \mathrm{EVP}(x)$ and  $\mathrm{EVP}(x) = \overline{v_{out}}$ with $x = x_e$.
	\end{enumerate}
	\label{th:SS}
\end{theorem}
\begin{IEEEproof}
	Please refer to Appendix D.
\end{IEEEproof}

Next, we consider the two scenarios in Theorem~\ref{th:SS} and obtain the closed-form solutions of $V(q_l, q_r)$ respectively. {We use a triple tuple $(V_l, x, C^{\infty})$ to denote the steady state, where $V_l$ and $x$ is derived from Theorem~\ref{th:SS}, and $C^{\infty}$ is calculated from (\ref{eq:simhjb}).}

\noindent{\bf 1) Computation Sufficient Scenario}

In this scenario, we consider the local task arrival rate $\overline{\lambda} < \mathrm{EVP}(x_e) + \frac{\overline{k}^2}{2c\beta}x_e$, and the remote computation capability is sufficient. Based on Theorem \ref{th:SS}, {the steady state $(V_{l,s}, V_{l,s}, C^{\infty})$} is
\begin{equation}
\overline{\lambda} = \mathrm{EVP}(V_{l,s})+\frac{\overline{k}^2}{2c\beta}V_{l,s}.
\end{equation}
Because the queue lengths are 0 in the steady state, based on (\ref{eq:cont_v}) and Theorem~\ref{th:HJB}, the optimal average cost can be denoted as
\begin{equation}
C^{\infty} = \beta \mathrm{EP}(V_{l,s}) + \frac{\overline{k}^2}{4c\beta}V^2_{l,s}.
\end{equation}

In the steady state, the arrival and departure rates are the same in a long-term sense. However, both the arrival and departure are time-varying, and the instantaneous arrival and departure rates are usually different. We define the difference between the instantaneous arrival and departure rates as follows:

\begin{definition}[{Dynamic Instantaneous Rate Estimation for Virtual Local Queue}]
	Denote $\epsilon$ as the instantaneous task rate difference between the input of virtual local queue and the output which includes transmission and local computation, i.e.,
	\begin{equation}
	\epsilon = \mathrm{EVP}(V_{l,c}) + \frac{\overline{k}^2}{2c\beta}V_{l,c} - \overline{\lambda},
	\label{eq:defeps}
	\end{equation}
	{where $V_{l,c}$ is the optimal value of $V_l$ under the instantaneous rate difference.}
	\hfill\IEEEQED
	\label{def:del}
\end{definition}
{According to Definition~\ref{def:del}, the corresponding average cost is }
\begin{equation}
C = \beta \mathrm{EP}(V_{l,c}) + \frac{\overline{k}^2}{4c\beta}V^2_{l,c}.
\end{equation}
Note that the instantaneous rate difference $\epsilon$ can be estimated by short-term statistics.

With the instantaneous state, we can solve the HJB equation (\ref{eq:simhjb}) with more accurate approximation and obtain the closed-form solution of $V(q_l, q_r)$.
\begin{theorem}[Asymptotic Closed-Form Priority Function in Computation Sufficient Scenario]
	
	For a given $\epsilon > 0$, {the priority function is expressed as}
	\begin{equation}
	V(q_l, q_r) = \frac{\alpha}{2\overline{\lambda}\epsilon}q^2_l + \frac{C-C^{\infty}}{\epsilon}q_l + \frac{\alpha}{2\overline{\lambda}\big[\overline{v_{out}} - BE_1\big(\frac{\beta N_0}{V_{l,c}B L}\big)\big]}q^2_r.
	\label{eq:closedform1}
	\end{equation}
	\label{th:closedform1}
\end{theorem}
\begin{IEEEproof}
	Please refer to Appendix E.
\end{IEEEproof}

The above theorem {considers} the solution in the case $\epsilon > 0$. When $\epsilon \leq 0$, we cannot solve the PDE in (\ref{eq:simhjb}) directly because the coefficient of $q_l$ in solution $V(q_l, q_r)$ is negative, which does not make sense for the physical meaning of the priority function. Instead, we try to find an approximation for the case $\epsilon \leq 0$.

To find an appropriate approximation of $\epsilon$, we need to consider the influence to $V(q_l, q_r)$ first. 
From (\ref{eq:closedform1}), the weight of $q_l$ is $\frac{\alpha}{2\overline{\lambda}\epsilon}$.
When $\epsilon > 0$, {the weight} of $q_l$ is a decreasing function of $\epsilon$ in $V(q_l, q_r)$. From Theorem~\ref{th:closedform1}, we derive the weight of $q_l$ tends to $+\infty$ when $\epsilon$ tends to 0. Based on the above analysis, we know that the weight of $q_l$ with $\epsilon \leq 0$ should be larger than that with $\epsilon > 0$, which means that the weight with $\epsilon \leq 0$  should be larger than $+\infty$. For a finite length queue, a sufficient large value of $\frac{1}{\epsilon}$ is enough to indicate the importance of $q_l$.
Thus, we approximate the difference $\epsilon$ to $\epsilon_0$ when $\epsilon \leq \epsilon_0$, where $\epsilon_0$ is a sufficient small constant under the condition $\epsilon_0 > 0$.

\begin{theorem}[Approximation Error for $\epsilon \leq 0$]
	The approximation error between the steady state $\frac{C - C^{\infty}}{\epsilon_0}$ and the optimal state $V_{l,s}$ is $O(\epsilon_0)$.
	\label{th:apprscincsc}
\end{theorem}
\begin{IEEEproof}
	Please refer to Appendix F.
\end{IEEEproof}

We summarize some insights from the optimal computation offloading with the closed-form virtual priority function as follows:
\begin{remark}[Insights in Computation Sufficient Scenario]
	From the closed-form priority function in (\ref{eq:closedform1}), we have 
	\begin{eqnarray}
	V_l &=& \frac{\alpha}{\overline{\lambda} \epsilon}q_l + \frac{C-C^{\infty}}{\epsilon},
	\label{eq:vl}\\
	V_r &=& \frac{\alpha}{\overline{\lambda}\big[\overline{v_{out}} - BE_1\big(\frac{\beta N_0}{V_{l,c}B L}\big)\big]}q_r.
	\end{eqnarray}
	From these expressions, we can extract the following insights:
	\begin{itemize}
		\item The weight of $q_l$ is a non-increasing function of $\epsilon$. With the same $q_l$, if the task rate difference of the local queue $\epsilon$ is small, our computation offloading policy will give a high power gain to reduce the local queue length.
		\item The local computation power $P_l$ is an increasing function of $q_l$, which is reasonable because a high task rate is required to reduce the local queue length when $q_l$ is large.
		\item If $q_r < \big(\frac{\alpha}{\overline{\lambda}\epsilon}q_l+\frac{C - C^{\infty}}{\epsilon} - \frac{\beta N_0}{BH}\big)\cdot\frac{\lambda \big(\overline{v_{out}} - BE_1(\frac{\beta N_0}{V_{l,c}BL})\big)}{\alpha}$, the transmission power $P_t$ is an increasing function of $q_l$ and a decreasing function of $q_r$. Otherwise, $P_t = 0$. It is not necessary to  push the computation tasks to the MEC server when $q_r$ is too large. With our policy, the local queue and the remote queue will keep in equilibrium until both of them be the steady state. 
		\hfill\IEEEQED
	\end{itemize}
	\label{coro:opcsincsc}
\end{remark}

\noindent{\bf 2) Computation Constrained Scenario}

In this scenario, we consider the local task arrival rate $\overline{\lambda} \geq \mathrm{EVP}(x_e) + \frac{\overline{k}^2}{2c\beta}x_e$, and the remote computation is constrained.
{We obtain the steady state $(V_{l,s}, x_s, C^{\infty})$} as follows:
\begin{equation}
\overline{\lambda} = \mathrm{EVP}(x_{s})+\frac{\overline{k}^2}{2c\beta}V_{l,s}
\end{equation}
\begin{equation}
\overline{v_{out}} = \mathrm{EVP}(x_{s})
\end{equation}
\begin{equation}
C^{\infty} = \beta \mathrm{EP}(x_{s}) + \frac{\overline{k}^2}{4c\beta}V^2_{l,s}
\end{equation}
Similar to Definition \ref{def:del}, we have the following definition for the remote task queue.
\begin{definition}[{Dynamic Instantaneous Rate Estimation for Virtual Remote Queue}]
	{Denote $\delta$ as the instantaneous task rate difference between the input and the output of virtual remote queue, i.e.,}
	\begin{equation}
	\delta = \overline{v_{out}} - \mathrm{EVP}(x_c),
	\label{eq:defdelta}
	\end{equation}
	{where $x_{c}$ is the optimal value of $x$ under the instantaneous rate difference.
		Combining the Definition \ref{def:del}, $V_{l,c}$ is determined by}
	\begin{equation}
	\epsilon = \mathrm{EVP}(x_c) + \frac{\overline{k}^2}{2c\beta}V_{l,c} - \overline{\lambda},
	\end{equation}
	\hfill\IEEEQED
	\label{def:der}
\end{definition}
{According to Definition~\ref{def:der}, the corresponding average cost is}
\begin{equation}
C = \beta \mathrm{EP}(x_c) + \frac{\overline{k}^2}{4c\beta}V^2_{l,c}
\end{equation}


With the instantaneous state, we can solve the HJB equation (\ref{eq:simhjb}) with more accurate approximation and obtain the closed-form solution of $V(q_l, q_r)$.
\begin{theorem}[Asymptotic Closed-Form Priority Function in Computation Constrained Scenario]
	
	For given $\epsilon > 0$ and $\delta > 0$, {the priority function is expressed as}
	\begin{equation}
	V(q_l, q_r) = \frac{\alpha}{2\overline{\lambda}\epsilon}q^2_l + \gamma\frac{C-C^{\infty}}{\epsilon}q_l + (1-\gamma)\frac{C-C^{\infty}}{\delta}q_r + \frac{\alpha}{2\overline{\lambda}\delta}q^2_r,
	\label{eq:closedform2}
	\end{equation}
	{where $\gamma = \big[\frac{(x_s + V_{l,s})\epsilon\delta}{2(\epsilon + \delta)(C-C^{\infty})} + \frac{V_{l,s}\epsilon^2}{2(\epsilon + \delta)(C-C^{\infty})} + \frac{\epsilon}{2(\epsilon + \delta)}\big]_\Gamma$, and $[\cdot]_\Gamma$ denotes the projection onto $\Gamma = \{\gamma:0 \leq \gamma \leq 1,\;x_c \leq \gamma \frac{C-C^{\infty}}{\epsilon} - (1-\gamma)\frac{C-C^{\infty}}{\delta} \leq x_s,\;V_{l,s} \leq \gamma \frac{C-C^{\infty}}{\epsilon} \leq V_{l,c}\}$.}
	\label{th:closedform2}
\end{theorem}
\begin{IEEEproof}
	Please refer to Appendix G.
\end{IEEEproof}

{For the cases with $\epsilon \leq 0$ or $\delta \leq 0$, similar to the computation sufficient scenario, we approximate the difference $\epsilon$ to $\epsilon_0$ when $\epsilon \leq \epsilon_0$, and the difference $\delta$ to $\delta_0$ when $\delta \leq \delta_0$, where $\epsilon_0$ and $\delta_0$ are sufficient small constants under the condition $\epsilon_0 >0$ and $\delta_0 > 0$. Using the similar approach with the proof of Theorem \ref{th:apprscincsc}, we obtain the following theorem:
	\begin{theorem}[Approximation Error for $\epsilon \leq 0$ or $\delta \leq 0$]
		The approximation error between the steady state $\gamma\frac{C - C^{\infty}}{\epsilon_0}$ and the optimal state $V_{l,s}$ is $O(\epsilon_0)$. Also, the approximation error between the steady state $\gamma\frac{C-C^{\infty}}{\epsilon_0} - (1-\gamma)\frac{C-C^{\infty}}{\delta_0}$ and the optimal state $x_s$ is $O(\delta_0)$.\hfill\IEEEQED
		\label{th:apprscinccc}
\end{theorem}}

Based on the closed-form solution $V(q_l, q_r)$ in Theorem \ref{th:closedform2}, we summarize the optimal computation offloading structure as follows:
{\begin{remark}[Insights in Computation Constrained Scenario]
		From the closed-form priority function in (\ref{eq:closedform2}), we have
		\begin{eqnarray}
		V_l &=& \frac{\alpha}{\overline{\lambda} \epsilon}q_l + \gamma\frac{C-C^{\infty}}{\epsilon},\\
		V_r &=& \frac{\alpha}{\overline{\lambda} \delta}q_r + (1 - \gamma)\frac{C-C^{\infty}}{\delta}.
		\end{eqnarray}
		From these expressions, we can extract the following insights:
		\begin{itemize}
			\item The weight of $q_l$ is a non-increasing function of $\epsilon$, which has the similar insights with those in computation sufficient scenario.
			\item The weight of $q_r$ is a non-increasing function of $\delta$. If the rate difference of the remote queue is large, our policy will reduce the influence of $q_r$ to the water level and increase the offloading rate of the local queue, which keeps the length of the remote queue to prevent the waste of computation resources. If the task rate difference is small, the policy will reduce the offloading for keeping the stability of the remote queue. 
			\item The local computation power $P_l$ is an increasing function of $q_l$, which has the similar insights with those in computation sufficient scenario.
			\item If $q_r < \big[\frac{\alpha}{\overline{\lambda}\epsilon}q_l+\gamma\frac{C - C^{\infty}}{\epsilon} - (1-\gamma)\frac{C-C^{\infty}}{\delta} - \frac{\beta N_0}{BH}\big]\cdot\frac{\lambda\delta}{\alpha}$, the transmission power $P_t$ is an increasing function of $q_l$ and a decreasing function of $q_r$. Otherwise, $P_t = 0$. 
			\hfill\IEEEQED
		\end{itemize}
		\label{coro:opcsinccc}
\end{remark}}

\subsection{Stability Conditions in Discrete Time System}
In this subsection, we show that the proposed offloading policy in Theorem~\ref{coro:ptstar} derived from the analysis in VCTS is also admissible in the original discrete time system.
{Specifically, we derive the following theorem to guarantee the system stability when using the computation offloading policy derived in Theorem~\ref{coro:ptstar} in the original discrete time system:}
\begin{theorem}[Stability in the Original Discrete-Time System]	\label{th:stable}
	Using the policy in Theorem~\ref{coro:ptstar} with the priority functions in Theorems~\ref{th:closedform1} and \ref{th:closedform2}, the local and remote queues in the original discrete time system are stable, i.e., $\lim_{n \to \infty}\mathbb{E}[Q^2_l(n)] < \infty$ and $\lim_{n \to \infty}\mathbb{E}[Q^2_r(n)] < \infty$.
\end{theorem}
\begin{IEEEproof}
	Please refer to Appendix H.
\end{IEEEproof}

{Now we obtain a delay-optimal computation offloading policy for origin Problem~\ref{prob:dopco}. In the next section, we will extend our policy to multiple MTs systems and derive some brief insights.}

{\section{Extension to Multi-MT Multi-server Scenarios}}
{In this section, we extend our derived computation offloading policy to multiple MTs MEC servers scenarios. Specifically, we list the main differences between the single-MT single-server scenario and the multi-MT multi-server scenario, and develop a multi-MT multi-server computation offloading policy by adopting learning approach.}

{\subsection{Main differences}}
{Consider a MEC system with $I$ MTs and $J$ MEC servers. Different from the single-MT single-server scenario, multiple MTs will share the limited wireless channel capability and the computation capabilities of the MEC servers in multi-MT multi-server scenario. Specifically, we need to consider
	\begin{itemize}
		\item \textbf{Conflict due to the limited communication capability}: For avoiding the communication collision, only a limited number of MTs are allowed to access the MEC servers and offload the computation tasks in each slot simultaneously. So in one slot, which MTs have the chances to access the MEC servers need to be determined.
		\item \textbf{Conflict due to the limited computation capability}: The remote queue lengths at the MEC servers influence the delay performance of all the MTs simultaneously. A computation task allocation should be determined between different MEC servers.
\end{itemize}}

The random task arrival, the local computation rate, transmission rate and the remote computation rate of MT $i$ and MEC server $j$ are formulated in the same manner as the single-MT single-server scenario in Section II. The notations used in multi-MT multi-server scenarios are summarized in Table~\ref{table:1}. We ignore the introductions of the notations which are extended from the single-MT single-server scenario for brevity.

\begin{table}
	\caption{List of Notations in Multi-MT Multi-Server Scenario}
	\centering
	\begin{tabular}{p{0.2\textwidth}<{\centering}p{0.6\textwidth}<{\centering}}
		\hline
		Parameter&Definition\\
		\hline
		$H_{i,j}(n)$&the local CSI from MT $i$ to MEC server $j$ in slot $n$\\
		\hline
		$\lambda_i(n)$&the random task arrival of MT $i$ in slot $n$\\
		\hline
		$v_{l,i}(P_{l,i}(n))$&the local computation rate and local computation power of MT $i$ in slot $n$\\
		\hline
		$v_{t,i,j}(P_{t,i,j}(n), H_{i,j}(n))$&the transmission rate and transmission power from MT $i$ to MEC server $j$ in slot $n$\\
		\hline
		$v_{out,j}(n)$&the computation rate of MEC server $j$\\
		\hline
		$\rho_{i,j}(n)$&the access indicator from MT $i$ to MEC server $j$\\
		\hline
	\end{tabular}
	\label{table:1}
\end{table}

{To avoid the conflict above, we denote $\rho_{i,j}(n)$ as the access indicator from MT $i$ to MEC server $j$ in slot $n$, where $\rho_{i,j}(n) = 1$ represents MT $i$ accessing MEC server $j$ successfully in slot $n$ and $\rho_{i,j}(n) = 0$ otherwise. Denote $\bm{\rho} = \{\rho_{i,j}(n)\}^{I,J}_{i,j=1}$ as the access solution for the MEC system. Because the wireless channel capacity is limited, we define $\bm{\rho_{f}}$ as the set of feasible access solutions for the MEC system, and a feasible access solution should satisfy $\bm{\rho}\in\bm{\rho_{f}}$.}

{The local task queue at MT $i$ is given by}
\begin{equation}
	Q_{l,i}(n+1) = [Q_{l,i}(n) - v_{l,i}(P_{l,i}(n))\tau - \sum_{j=1}^{J}\rho_{i,j}(n)v_{t,i,j}(P_{t,i,j}(n), H_{i,j}(n))\tau]^+ + \lambda_i(n)\tau,
\end{equation}
{and the remote task queue at the MEC server $j$ is given by}
\begin{equation}
	Q_{r,j}(n+1) = [Q_{r,j}(n) - v_{out,j}(n)\tau]^+ + \sum_{i=1}^{I}\rho_{i,j}(n)v_{t,i,j}(P_{t,i,j}(n), H_{i,j}(n))\tau.
\end{equation}

{Denote $\{Q_{l,i}(n)\}^{I}_{i=1}$ and $\{Q_{r,j}(n)\}^{J}_{j=1}$ as the set of local queue lengths and the set of remote queue lengths. Also, denote	$\{H_{i,j}(n)\}^{I,J}_{i,j=1}$ as the set of the local CSI of MT $i$ for accessing MEC server $j$. In multi-MT multi-server scenario, the computation offloading policy $\Omega(S(n))$ includes the access solutions $\bm{\rho}$, which are determined at the MEC servers. Then, based on the multi-MT multi-server queue dynamics, we can rewrite the transition probability of the random process $S(n) = (\{Q_{l,i}(n)\}^{I}_{i=1}, \{Q_r(n)\}^{J}_{j=1}, \{H_{i,j}(n)\}^{I,J}_{i,j=1})$ as}
\begin{equation}
	\begin{aligned}
	\mathrm{Pr}&\big[S(n+1)\mid S(n), \Omega(S(n))\big]\\
	=&\mathrm{Pr}\big[\{H_{i,j}(n+1)\}^{I,J}_{i,j=1}\big]\cdot \mathrm{Pr}\big[\{Q_{l,i}(n+1)\}^{I}_{i=1} \mid \{Q_{l,i}(n)\}^{I}_{i=1}, \{H_{i,j}(n)\}^{I,J}_{i,j=1}, \Omega(S(n))\big] \\
	&\cdot \mathrm{Pr}\big[\{Q_{r,j}(n+1)\}^{J}_{j=1} \mid \{Q_{l,i}(n)\}^{I}_{i=1}, \{Q_{r,j}(n)\}^{J}_{j=1}, \{H_{i,j}(n)\}^{I,J}_{i,j=1}, \Omega(S(n))\big],\\
	=&\mathrm{Pr}\big[\{H_{i,j}(n+1)\}^{I,J}_{i,j=1}\big]\cdot\prod_{i=1}^{I}\mathrm{Pr}(\lambda_i(n))\cdot\prod^{J}_{j=1}\mathrm{Pr}(v_{out,j}(n)).\nonumber
	\end{aligned}
\end{equation}

{From the problem formulation aspect, the average delay $\overline{D(\Omega)}$ and the average power consumption $\overline{P(\Omega)}$ starting from a given initial state $S(0)$ are given by}
\begin{equation}
	\overline{D_i(\Omega)} = \limsup_{N\to +\infty}\frac{1}{N}\sum_{n = 0}^{N-1}\mathbb{E}^{\Omega}\bigg[\frac{1}{\overline{\lambda_i}}\big(Q_{l,i}(n) + \sum^J_{j=1}Q_{r,j}(n)\big)\bigg],
	\label{eq:newd}
\end{equation}

\begin{equation}
	\overline{P_i(\Omega)} = \limsup_{N\to +\infty}\frac{1}{N}\sum_{n = 0}^{N-1}\mathbb{E}^{\Omega}\big(P_{l,i}(n) + \sum^J_{j=1}\rho_{i,j}(n)P_{t,i,j}(n)\big).
\end{equation}
{Based on Theorem~\ref{th:SCO} and Theorem~\ref{th:HJB}, we have the following HJB equation for multiple MTs scenarios to derive the priority function:}
\begin{equation}
	\begin{aligned}
	&\min\limits_{\{P_{l,i}\}, \{P_{t,i}\}, \{\rho_{i,j}\}} \sum^{I}_{i=1}\sum^{J}_{j=1}\bigg\{\mathbb{E}\bigg[\frac{\alpha}{\overline{\lambda_i}}\big(q_{l,i} + \sum^J_{j'=1}q_{r,j'}\big) + \beta\big(P_{l,i} + P_{t,i,j}\big) + \frac{\partial V(q_{l,i}, q_{r,j})}{\partial q_{l,i}}\big[\overline{\lambda_i} \\
	&- v_{l,i}(P_{l,i})- \rho_{i,j}v_{t,i,j}(P_{t,i,j}, H_{i,j})\big] + \frac{\partial V(q_{l,i}, q_{r,j})}{\partial q_{r,j}}\big[\rho_{i,j}v_{t,i,j}(P_{t,i,j}, H_{i,j}) - \overline{v_{out,j}}\big]\bigg]- C_{i,j}^{\infty}\bigg\} = 0.
	\end{aligned}
\end{equation}

{\subsection{Learning-based Computation Offloading policy}}
{For deriving the priority function of each MT in multi-MT multi-server scenario, the expectations of several variables need to be calculated like Theorem~\ref{th:simhjb}. Specifically, the transmission power and the transmission rate are different from $\mathrm{EP}(x)$ and $\mathrm{EVP}(x)$ because of the access indicator $\rho_{i,j}$. Here we adopt a learning approach~\cite{YR2017} to obtain $\mathrm{EP}_{i,j}(x) = \mathbb{E}[\rho_{i,j}P_{t,i,j}] = \widehat{\rho_{i,j}}\mathrm{EP}(x)$ and $\mathrm{EVP}_{i,j}(x) = \mathbb{E}[\rho_{i,j}v_{t,i,j}(P_{t,i,j},H_{i,j})] = \widetilde{\rho_{i,j}}\mathrm{EVP}(x)$ based on the historical statistic informations, where $\widehat{\rho_{i,j}}$ and $\widetilde{\rho_{i,j}}$ introduce the effects of the average access ratio to the transmission power and the transmission rate respectively. With the same approach in Section~V, we derive the priority function for each MT as}
\begin{equation}
	V(q_{l,i}, q_{r,j})=
	\left\{
	\begin{aligned}
	&\frac{\alpha}{2\overline{\lambda_i}\epsilon_i}q^2_{l,i} + \frac{C_{i,j}-C^{\infty}_{i,j}}{\epsilon_i}q_{l,i} + \frac{\alpha}{2\overline{\lambda_i}\big[\overline{v_{out,j}} - \widetilde{\rho_{i,j}}BE_1\big(\frac{\beta N_0}{V_{l,c,i}B L_i}\big)\big]}q^2_{r,j}, &(a)\\
	& \frac{\alpha}{2\overline{\lambda_i}\epsilon_i}q^2_{l,i} + \gamma_i\frac{C_{i,j}-C^{\infty}_{i,j}}{\epsilon_i}q_{l,i} + (1-\gamma_i)\frac{C_{i,j}-C^{\infty}_{i,j}}{\delta_{i,j}}q_{r,j} + \frac{\alpha}{2\overline{\lambda_i}\delta_{i,j}}q^2_{r,j}, &(b)
	\end{aligned}
	\right.
\end{equation}
{in computation sufficient scenario $(a)$ and in computation constrained scenario $(b)$, where $C^{\infty}_{i,j} = \beta \widehat{\rho_{i,j}}\mathrm{EP}(x_{s,i,j}) + \frac{\overline{k_i}^2}{4c_i\beta}V^2_{l,s,i}$ and $C_{i,j} = \beta \widehat{\rho_{i,j}}\mathrm{EP}(x_{c,i,j}) + \frac{\overline{k_i}^2}{4c_i\beta}V^2_{l,c,i}$.}

{Based on the priority functions above, the delay-optimal computation offloading policy $\Omega(S) = (\{P^*_{l,i}\}, \{P^*_{t,i,j}\}, \{\rho^*_{i,j}\})$can be expressed as}
\begin{eqnarray}
	&P^*_{l,i} = \frac{\overline{k_i}^2}{4c_i\beta^2}V^2_{l,i},\\
	&P^*_{t,i,j} = \rho^*_{i,j}\bigg(\frac{B}{\beta}(V_{l,i} - V_{r,j}) - \frac{N_0}{H_{i,j}}\bigg)^+,\\
	&\{\rho^*_{i,j}\} = \arg\min\limits_{\{\rho_{i,j}\}}\sum_{i=1}^{I}\bigg[\alpha\overline{D_i(\Omega)} + \beta\overline{P_i(\Omega)} + V(Q_{l,i}, Q_{r,j}) \bigg],
	\label{eq:rstar}
\end{eqnarray}
{where $P^*_{l,i}$ and $P^*_{t,i,j}$ are determined at each MT $i$, and the access solutions $\{\rho^*_{i,j}\}$ are determined at the MEC servers. In each slot, each MT first achieves the optimal $P^*_{l,i}$ and $P^*_{t,i,j}$ under $\rho^*_{i,j}=1$, then the MEC servers\footnote{When the communication time between different MEC servers and different MTs is non-negligible, the proposed policy can still be applied by adding $\tau_{max}$ in (\ref{eq:newd}), where $\tau_{max}$ denotes the maximal communication delay for exchange the control signals.} collect the power allocation policies of all the MTs and derive the access solutions by (\ref{eq:rstar}). After that, the actual offloading processes will be executed and only the selected MTs have the opportunities to offload the computation tasks.}

\section{Performance Evaluation}
{In this section, we investigate the characteristics and evaluate the performance of the proposed computation offloading policy by simulation. First, we analyze the characteristics of the proposed policy. Second, we compare the performance of the proposed policy with several conventional approaches in not only {the single-MT single-server scenario but also the multi-MT multi-server scenario.} Finally, we summary the insights extracted from the simulation results.}

\subsection{{Performance Analysis}}
{We first investigate the tradeoff between the delay and the power, the performance impact of the imperfect CSI, and the performance gap between the closed-form computation offloading policy and the relative learning-based policy.}

\begin{figure*}
	\centering
	\subfigure[Influence of performance tradeoff]{
		\includegraphics[width=0.33\textwidth]{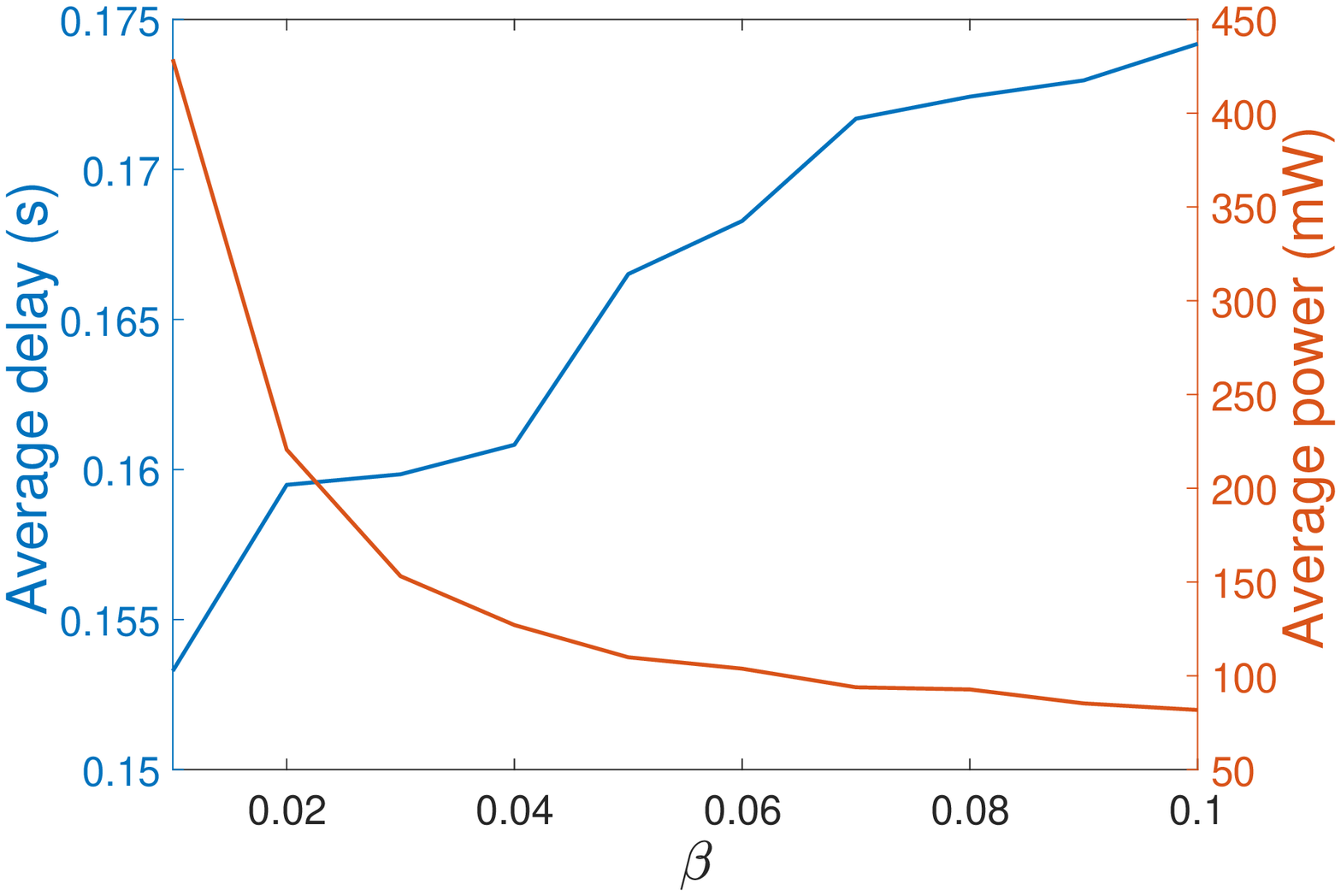}
		\label{fig:beta}
	}\hspace{-15pt}
	\subfigure[Influence of imperfect CSI]{
		\includegraphics[width=0.33\textwidth]{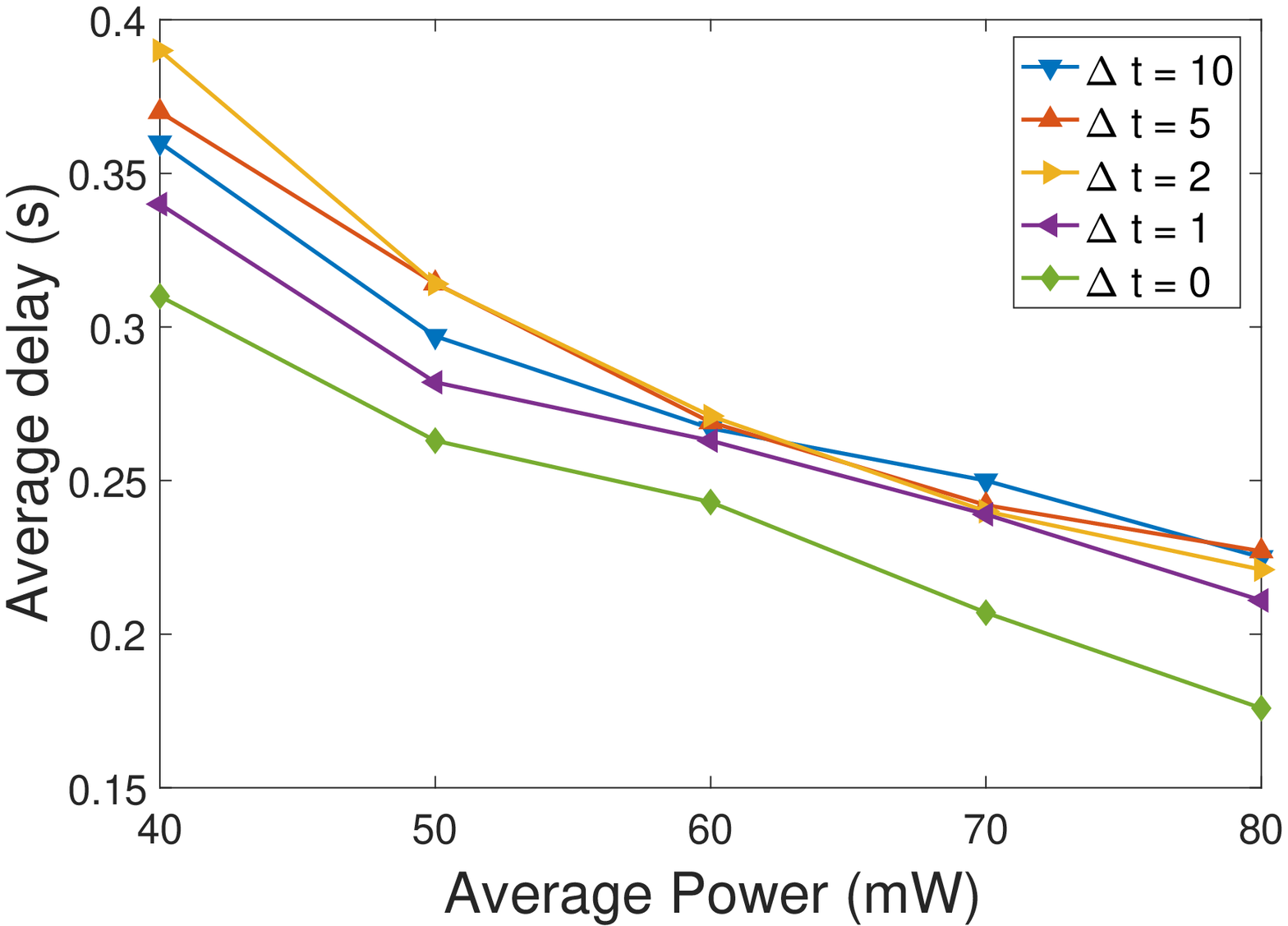}
		\label{fig:csi}
	}\hspace{-15pt}
	\subfigure[Influence of priority functions]{
		\includegraphics[width=0.33\textwidth]{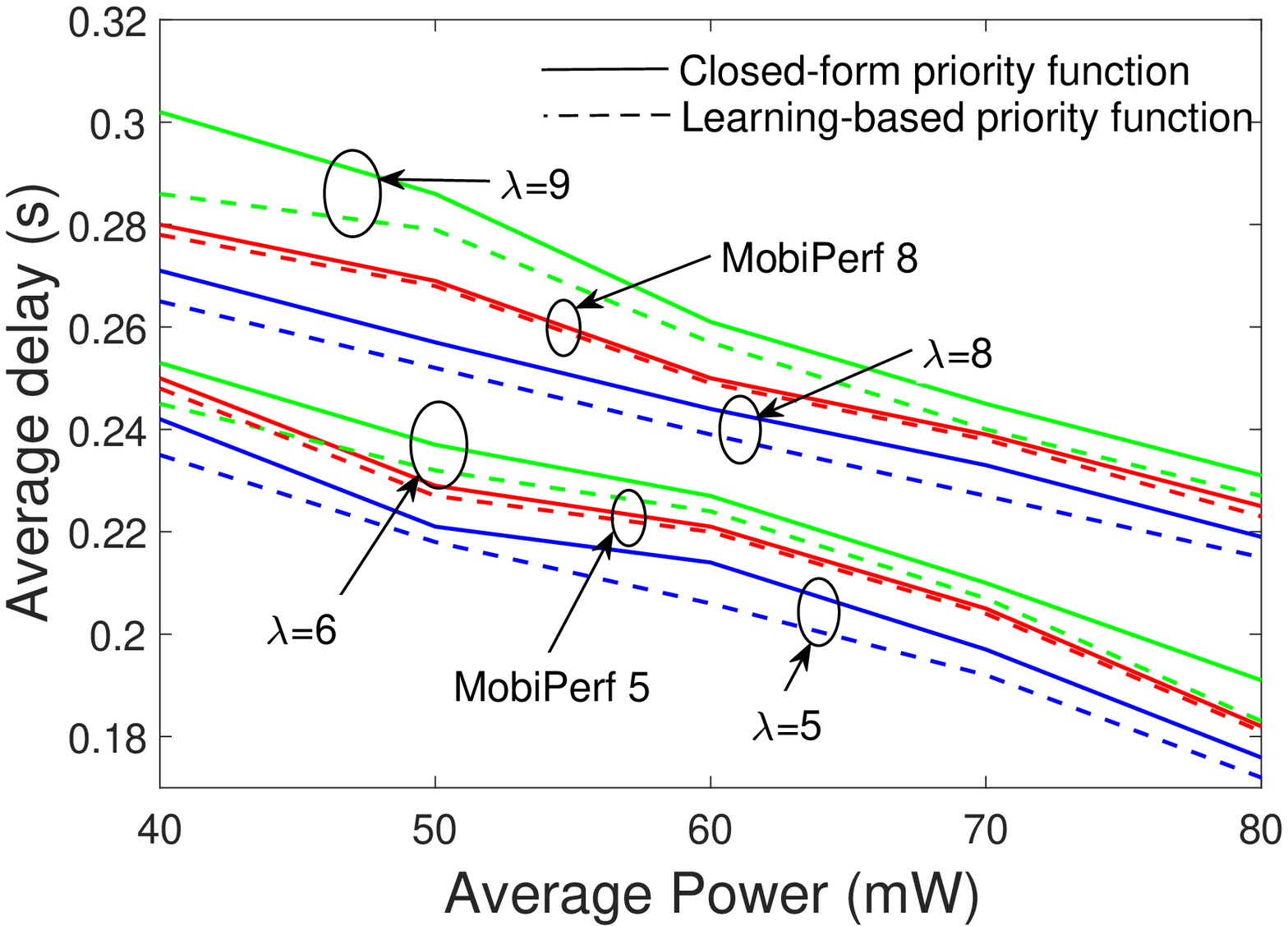}
		\label{fig:vfunc}
	}
	\caption{Performance Analysis}
\end{figure*}

{Fig.~\ref{fig:beta} shows the delay performance and the power consumption versus the weight $\beta$. It can be observed that the average delay increases and the average power consumption decreases with the increase of $\beta$. In particular, the average power grows rapidly when $\beta$ becomes small, because the average power increases exponentially with the increase of the transmission rate and increases squarely with the increase of the local computation rate.}

{Fig.~\ref{fig:csi} depicts the delay performance under different imperfect CSIs. We use the historical CSIs in different previous time to indicate the imperfection of the CSI and denote $\Delta t$ as the time interval (per slot) between the current system state and the CSI state. {It can be observed that the performance degradation exists but is not significant. In particular, there is not an obvious trend with the increase of $\Delta t$, which is because the CSI is i.i.d. over slots. In practical scenarios, our policy can achieve better delay performance due to the time-correlated CSI.}}

{Fig.~\ref{fig:vfunc} demonstrates the delay performance based on closed-form computation offloading policy through comparing with the relative learning based policy. The asymptotic performance is resulted from the uncertain rate estimation, so we compare the delay performance based on different random task arrivals for achieving some insights. MobiPerf~5 and MobiPerf~8 are two random task arrivals based on an open dataset Mobile Open Data by MobiPerf~\cite{opensource}, which average arrival rates are $5.54$ packet/s and $8.46$ packets/s, respectively. Under the same environment conditions except for the average arrival rate, the six curves below are in computation sufficient scenario, and the six curves above are in computation constrained scenario. It can be observed that the delay performance gaps between the closed-form computation offloading policy and the relative learning based policy are limited in 0.0018s/packet (The maximal gap is 0.0162s when $\lambda$ = 9 packet/s). Especially, the gaps based on the real data are smaller than other random arrivals. It is because the rate variation of the real task arrival is smaller than that of the random data, because of the continuity of real user usage.}

\subsection{{Performance Comparison}}
{We evaluate the performance of the proposed closed-form delay-optimal computation offloading policy, and compare it with the following five baselines:}
\begin{itemize}
		\item \textbf{Baseline 1 -- Greedy Throughput-Optimal Offloading (GT)}~\cite{VS2010}: The sum of the local computation power and the computation offloading power is fixed in each slot.
		\item \textbf{Baseline 2 -- CSI-Only Water-Filling Offloading (COWF)}~\cite{VS2010}: The local computation power is fixed, and the computation offloading policy depends on the CSI only. 
		\item \textbf{Baseline 3 -- Queue-Weighted Water-Filling Offloading (QWWF)}~\cite{LH2013}: The local computation power is fixed, and the computation offloading policy depends on the CSI and the LQSI~\cite{MA2004}.
		\item \textbf{Baseline 4 -- Lyapunov optimization-based dynamic computation offloading (LODCO)}~\cite{YM2016}: Both the local computation power and the computation offloading power are derived based on Lyapunov optimization. The computation task which cannot be executed in deadline will be dropped.
		\item \textbf{Baseline 5 -- Task Scheduling-based Offloading (TSO)}~\cite{EG1-JL2016}: The delay-optimal task scheduling strategy is derived by an one-dimensional search algorithm. The power is transformed from the task scheduling decision.
\end{itemize}

{Also, the simulation settings are as follows unless other wise stated.
The carrier sensing distance $d$ is $100$m, and the path gain is calculated as $L = 15.3 + 37.6\log_{10}d$ with the fading coefficient distributed as $\mathcal{CN}(0, 1)$. The system bandwidth is $10$MHz and the additive Gaussian noise power is $N_0 = -174\mathrm{dBm/Hz}$. The arrival of computation tasks are the the Mobile Open Data by MobiPerf and random arrivals with mean $\overline{\lambda}$ (packet/s). The computation parameter is set as $\overline{k} = 10^{-7}$ and $c = 3.5 * 10 ^{-12}$. For comparison, the delay performances of different baselines are evaluated with the same average power by adjusting $\beta$. We run the simulation for 100 times to obtain the average performance, and consider 500 time slots whose duration $\tau$ is $0.1$s. We compare the performance in single-MT single-server scenario and multi-MT multi-server scenario, respectively.}

{\noindent{\bf 1) Single-MT Single-Server Scenario}}

\begin{figure*}
	\centering
	\subfigure[Task arrival rate in computation sufficient scenario]{
		\includegraphics[width=0.47\textwidth]{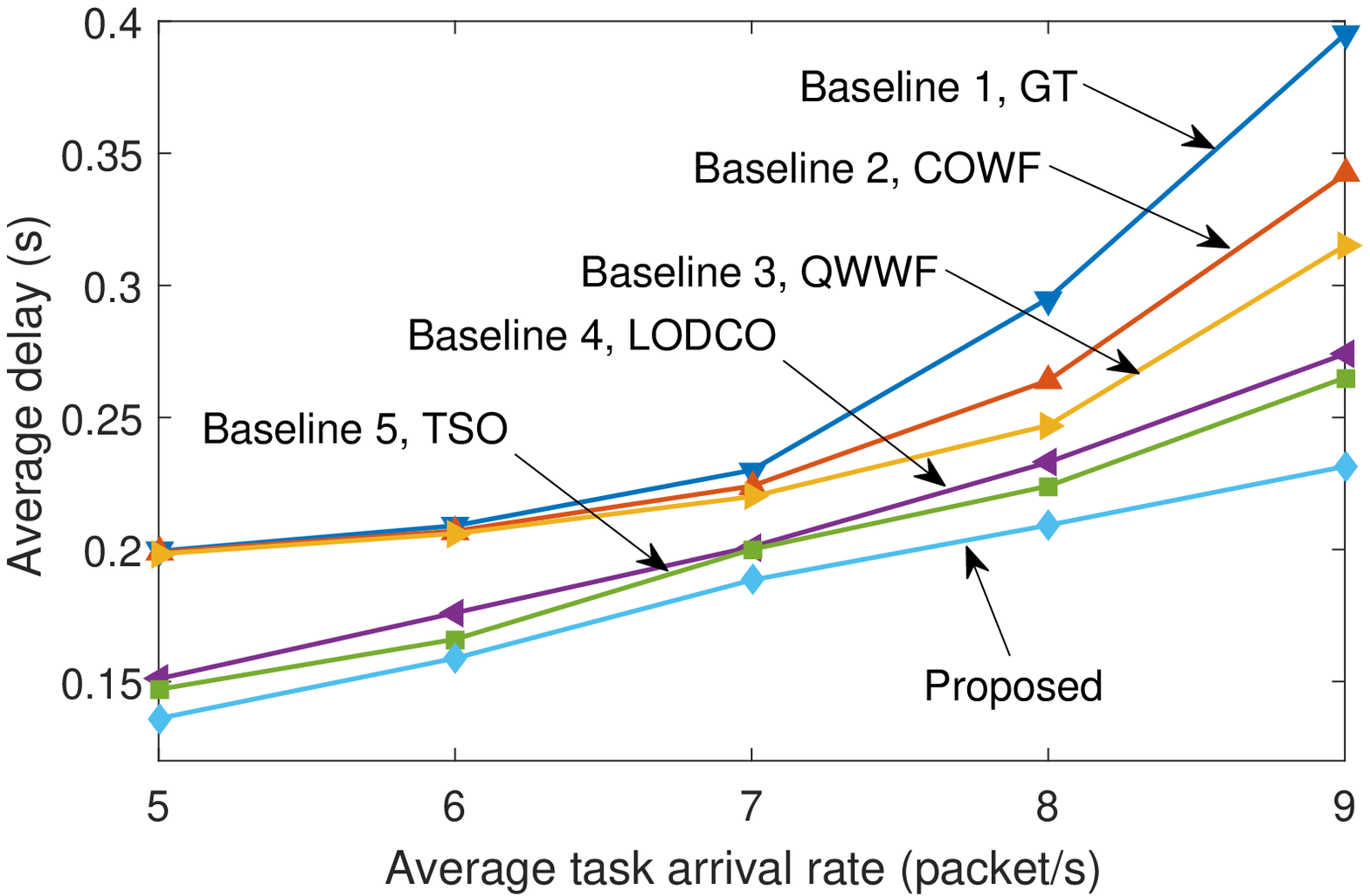}
		\label{fig:csc_lambda}
	}
	\subfigure[Task arrival rate in computation constrained scenario]{
		\includegraphics[width=0.47\textwidth]{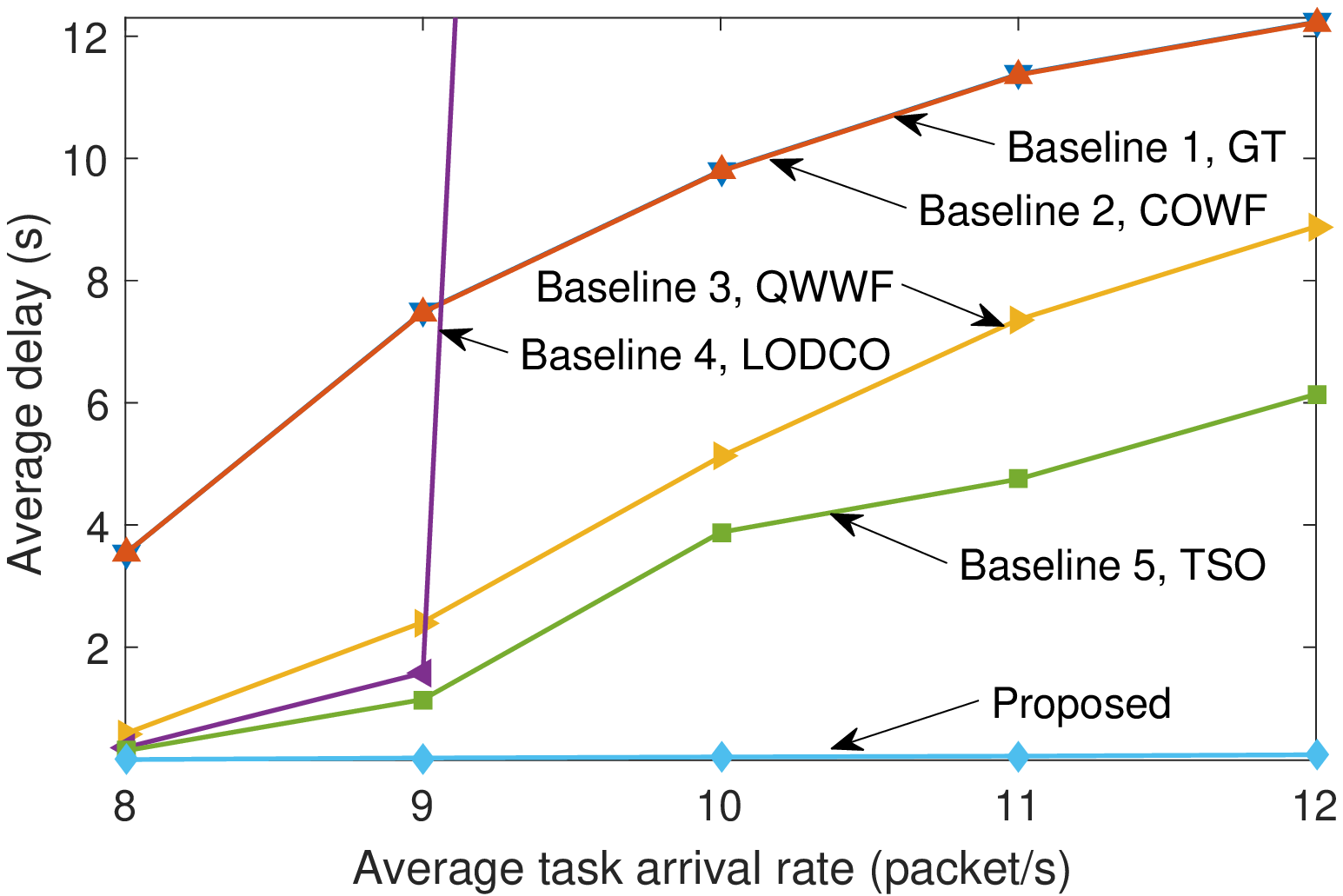}
		\label{fig:ccc_lambda}
	}
	
	\subfigure[Power consumption in computation sufficient scenario]{
		\includegraphics[width=0.47\textwidth]{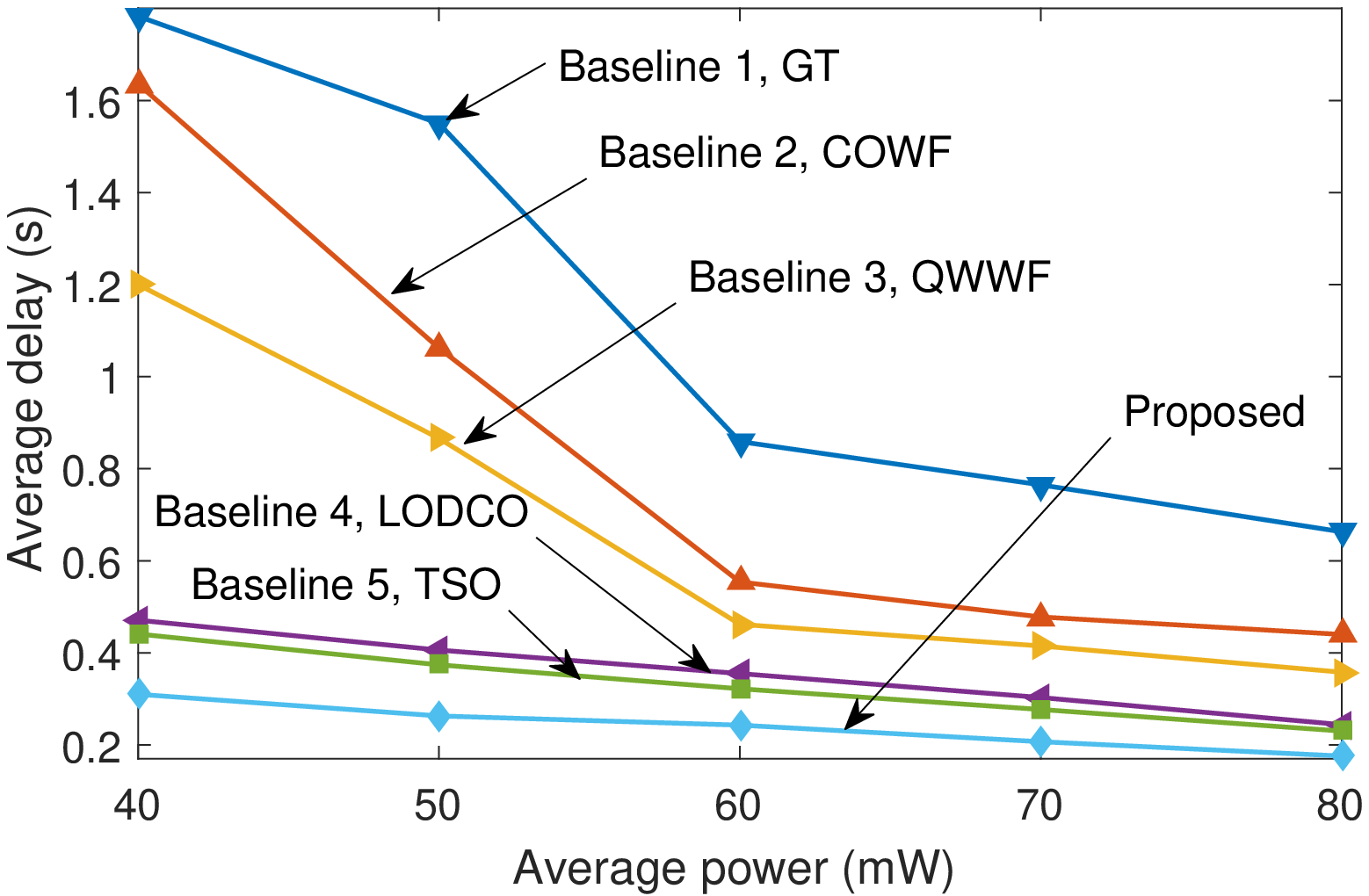}
		\label{fig:csc_power}
	}
	\subfigure[Power consumption in computation constrained scenario]{
		\includegraphics[width=0.47\textwidth]{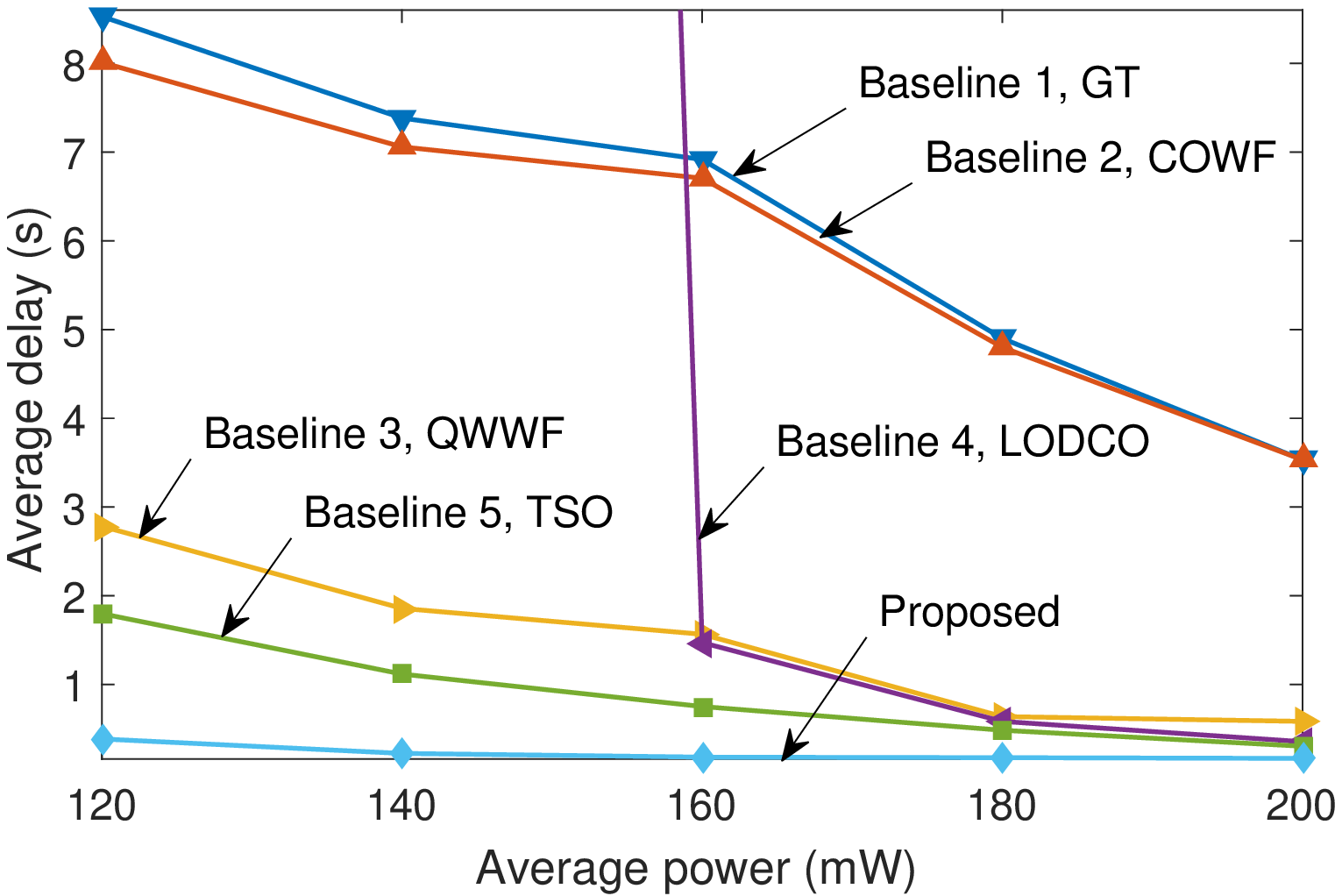}
		\label{fig:ccc_power}
	}
	\caption{Delay performance with different environment conditions}
	\label{fig:ec}
\end{figure*}

{We investigate the delay performance with different environment conditions and different computation capabilities, respectively. First, we investigate the delay performance with different environment conditions.}
{Fig.~\ref{fig:csc_lambda} and Fig.~\ref{fig:ccc_lambda} show the delay performance versus different task arrival rates. The average power is set to $100$ mW in computation sufficient scenario and $200$ mW in computation constrained scenario. The proposed policy achieves significant performance gain over all the baselines. Especially in computation-constrained scenario, the large performance gap indicates the importance of the RQSI. As the average task arrival rate increases, the backlog of the remote queue in other baseline increases and the performance gap increases. The proposed policy maintains the backlog of the remote queue to guarantee the offloading performance in a significative region.}

{Fig.~\ref{fig:csc_power} and Fig.~\ref{fig:ccc_power} illustrate the delay performance versus different average powers. The $\overline{\lambda}$ is set to $5$ packet/s in computation sufficient scenario and $8$ packet/s in computation constrained scenario. The proposed policy is not sensitive to the average power compared with other baselines. It is because the proposed policy can adjust the transmission power through the multi-level water-filling structure based on the CSI, the LQSI and the RQSI. For only considering the proposed policy, we can achieve an excellent tradeoff between delay and power by adjusting the weight $\beta$, which has discussed in the previous subsection.}

\begin{figure*}
	\centering
	\subfigure[Local computation rate in computation sufficient scenario]{
		\includegraphics[width=0.47\textwidth]{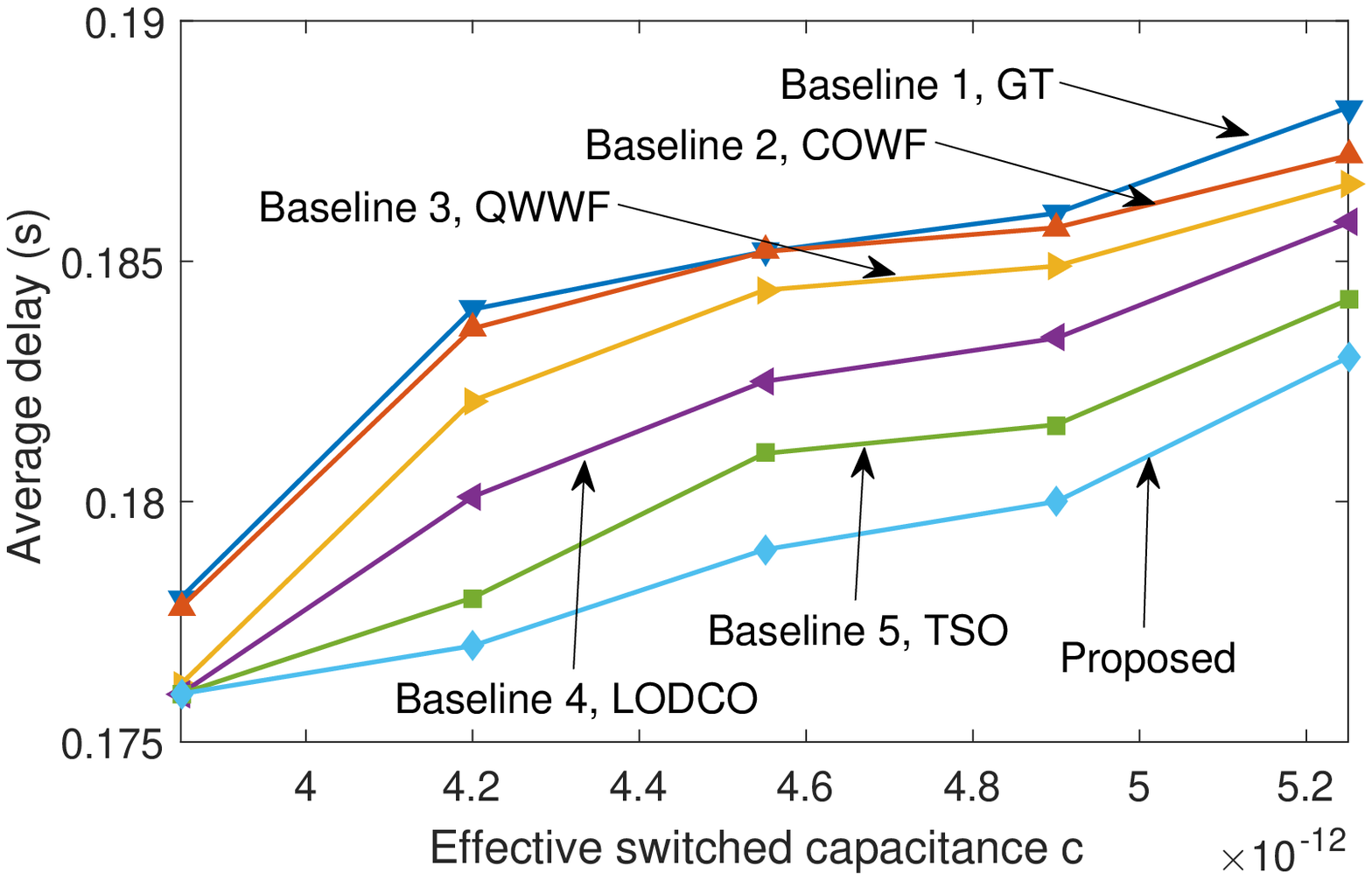}
		\label{fig:csc_c}
	}
	\subfigure[Local computation rate in computation constrained scenario]{
		\includegraphics[width=0.47\textwidth]{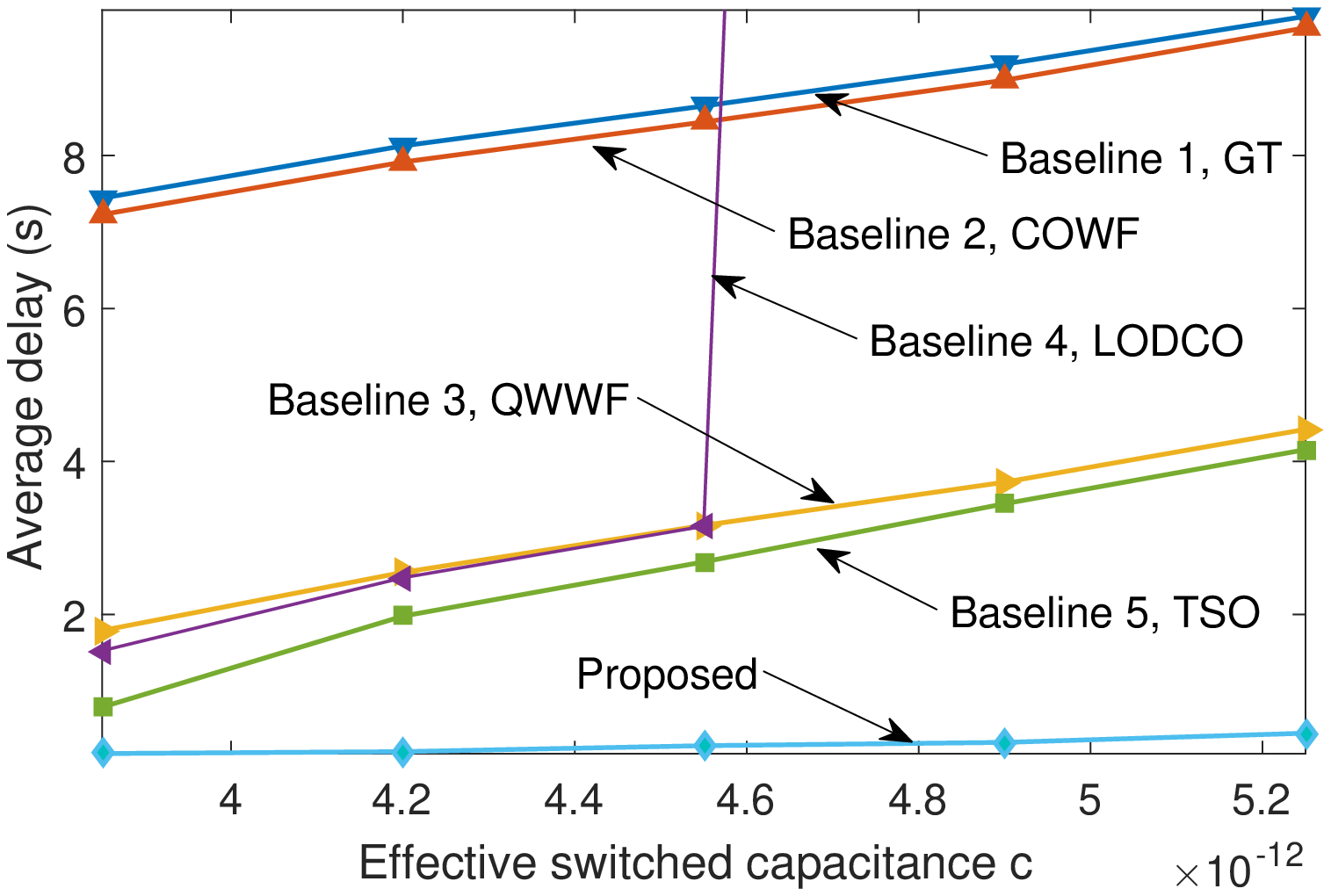}
		\label{fig:ccc_c}
	}
	\subfigure[Remote computation rate in computation sufficient scenario]{
		\includegraphics[width=0.47\textwidth]{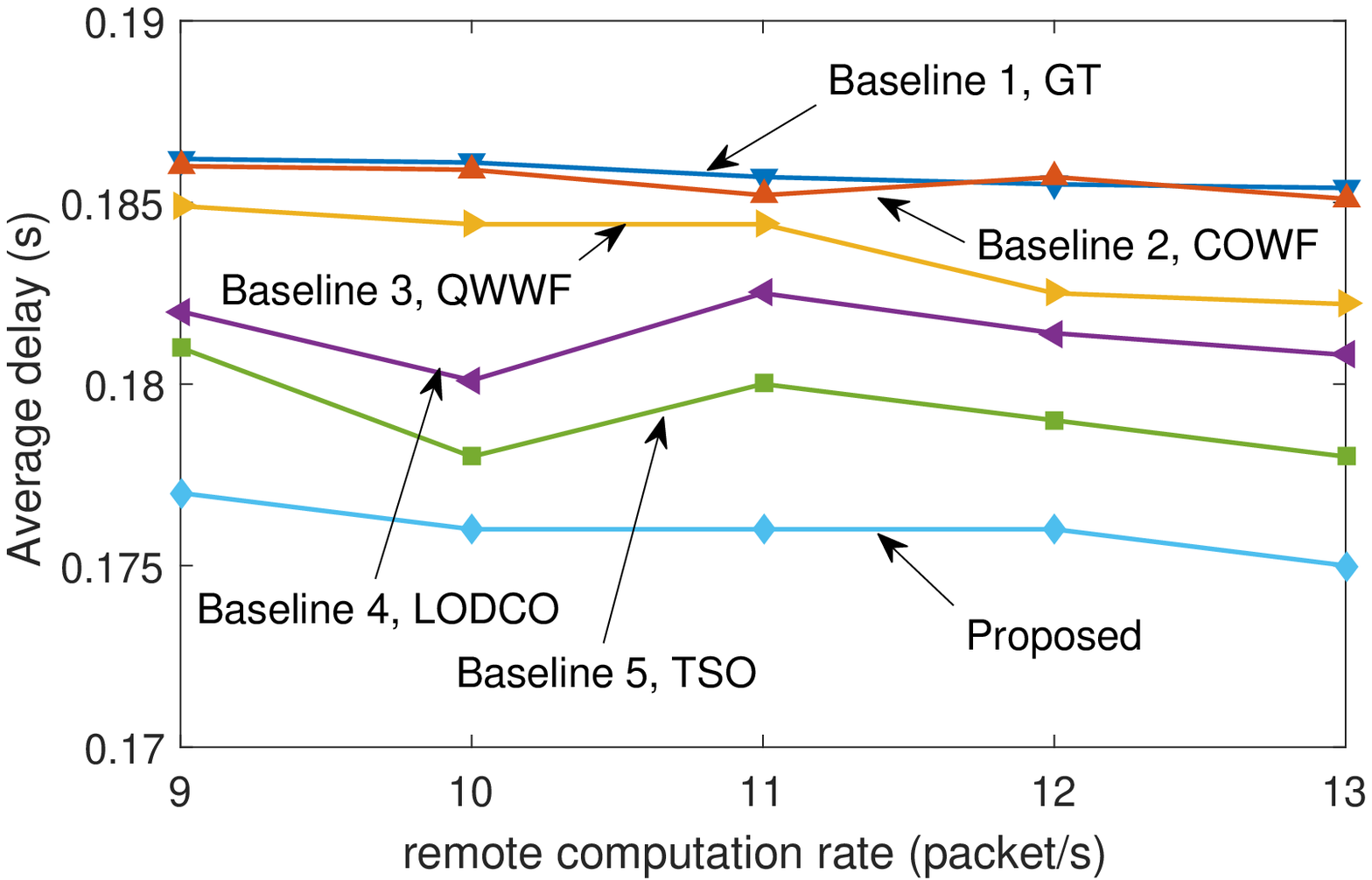}
		\label{fig:csc_vout}
	}
	\subfigure[Remote computation rate in computation constrained scenario]{
		\includegraphics[width=0.47\textwidth]{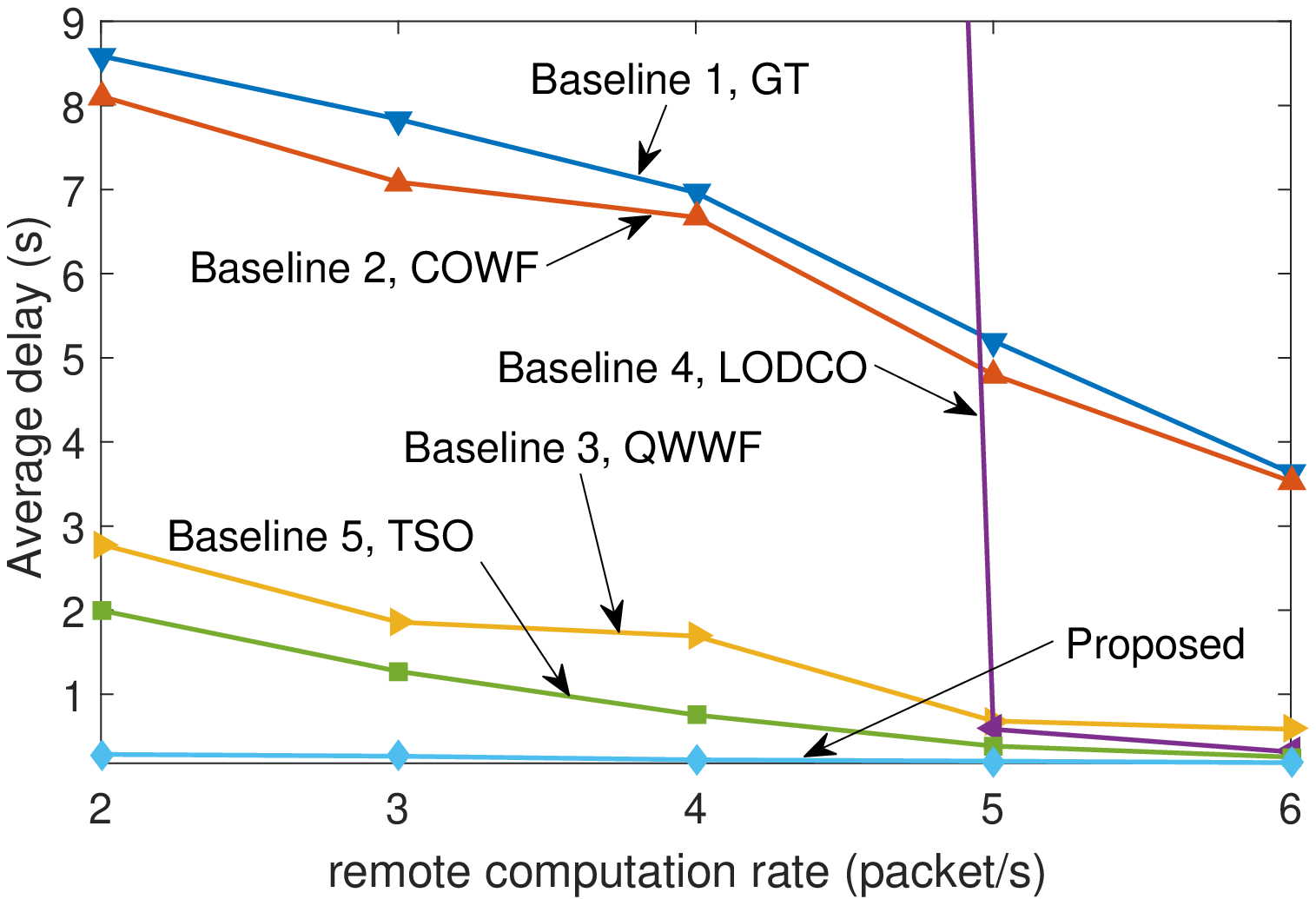}
		\label{fig:ccc_vout}
	}
	\caption{Delay performance with different computation capabilities}
	\label{fig:cc}
\end{figure*}

{Next, we reveal the delay performance with different computation capabilities.}
{Fig.~\ref{fig:csc_c} and Fig.~\ref{fig:ccc_c} depict the delay performance versus different local computation capabilities. With the increase of $c$, the local computation capability decreases under the same power constraint (based on~(\ref{vl2})), and the average delay increases. Note that the performance degradation in computation constrained scenario is larger than that in computation sufficient scenario. It means that the proposed policy is more sensitive to the local computation capability in computation constrained scenario. This result is in accordance with the practical situation that more computation tasks will be computed locally in computation constrained scenario.}

{Fig.~\ref{fig:csc_vout} and Fig.~\ref{fig:ccc_vout} present the delay performance versus different remote computation capabilities. With the increase of $\overline{v_{out}}$, the remote computation capability increases, the average delay in computation sufficient scenario is almost same, and the average delay in computation constrained scenario decreases. In computation sufficient scenario, the backlog of the remote queue is always zero, which means the influence of the remote queue on the average delay is very limited. This characteristic leads to the results in computation sufficient scenario. In computation constrained scenario, the average delay decreases because the influence of the remote queue on the average delay is remarkable.}

\begin{figure*}
	\centering
	\subfigure[Number of MT in computation sufficient scenario]{
		\includegraphics[width=0.47\textwidth]{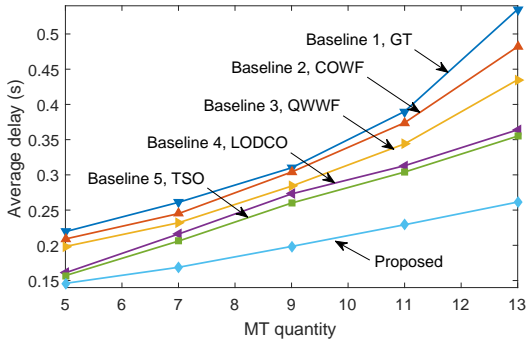}
		\label{fig:csc_mt}
	}
	\subfigure[Number of MT in computation constrained scenario]{
		\includegraphics[width=0.47\textwidth]{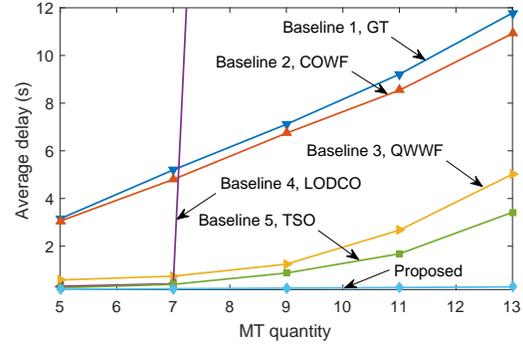}
		\label{fig:ccc_mt}
	}
	\subfigure[Number of server in computation sufficient scenario]{
		\includegraphics[width=0.47\textwidth]{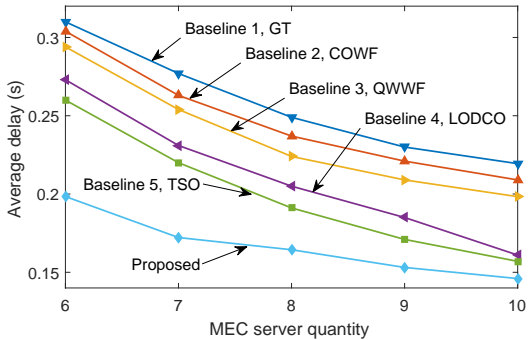}
		\label{fig:csc_mec}
	}
	\subfigure[Number of server in computation constrained scenario]{
		\includegraphics[width=0.47\textwidth]{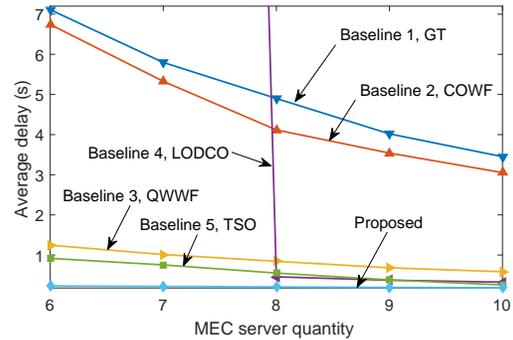}
		\label{fig:ccc_mec}
	}
	\caption{Delay performance with different numbers of MTs and MEC servers}
	\label{fig:multi}
\end{figure*}

{\noindent{\bf 2) Multi-MT Multi-Server Scenario}}

{We consider only one MT is allowed to access each MEC server in each slot in the following test. Fig.~\ref{fig:csc_mt} and Fig.~\ref{fig:ccc_mt} indicate the delay performance with different numbers of MTs.  The average task arrival rates of MTs are $5$ packet/s. The average computation rates of MEC servers are $13$ packet/s/MT in computation sufficient scenario, and are $3$ packet/s/MT in computation constrained scenario. We set $5$ MEC servers. With the increase of the number of MTs, the average delay increases.}
{Fig.~\ref{fig:csc_mec} and Fig.~\ref{fig:ccc_mec} indicate the delay performance with different numbers of MEC servers. We set the parameters with the same manner as the above simulation. We set $10$ MTs both in two tests. With the increase of the number of MEC servers, the average delay decreases. This result is similar with that in the previous simulation. Because only a limited number of MTs are allowed to access the MEC servers, the local queue backlogs of MTs which cannot offload the tasks lead to the delay performance degradation.}

\subsection{Summary}
{We investigate the characteristics and evaluate the performance of the proposed policy in this section. In the performance analysis, we find that we can adjust the tradeoff between delay and power by adjusting the parameter $\beta$. Also, we indicate the limited performance degradation of the proposed policy with imperfect CSI and demonstrate the closed-form policy is very close to the learning based real optimal policy. In the performance comparison, we compare the proposed policy with other conventional approaches under different environment conditions and computation capabilities, both in single-MT single-server scenario and multi-MT multi-server scenario. Several insights are obtained from the simulation results. All the analysis and comparisons show the proposed policy achieves excellent performance in computation constrained MEC systems.}

\section{Conclusion}
In this paper, we construct a MDP framework to optimize the delay performance in computation-constrained MEC systems. To address the technical challenges, we proceed by three steps. First, we reformulate the optimization problem using an infinite horizon average cost MDP, and then adopt a VCTS with reflections to deal with the curse of dimensionality. Second, we derive the closed-form approximate priority functions for the MDP by dynamic estimation of instantaneous rates in both the computation sufficient and computation constrained scenarios. Third, we propose a closed-form multi-level water-filling computation offloading solution to characterize the influence of not only the LQSI but also the RQSI. After that, we extend our policy into multi-MT multi-server scenario. Finally, we analyze the delay performance and the simulation results show that the proposed computation offloading scheme outperforms the conventional approaches.


\section*{Appendix A: Proof of Theorem \ref{th:SCO}}
Based on \emph{Proposition 4.6.1} of \cite{DP2005}, the sufficient conditions for the optimality of Problem \ref{prob:dopco} are that there exists a $(\theta^*, \{V^*(S)\})$ which satisfies the following Bellman equation and $V^*$ satisfies the transversality condition in (\ref{eq:trnasver}) for all the admissible offloading policy $\Omega$ and initial state $S(0)$:
\begin{equation}
	\begin{aligned}
	\theta^* + V^*(S) &= \min\limits_{\Omega(S)}\bigg[g(n)+\sum_{S'}\mathrm{Pr}\big[S'\mid S, \Omega(S)\big]V^*(S')\bigg]\\
	&=\min\limits_{\Omega(S)}\bigg[g(n)+\sum_{Q'_l, Q'_r}\sum_{H'}\mathrm{Pr}\big[Q'_l, Q'_r\mid Q_l, Q_r, \Omega(S)\big]\mathrm{Pr}[H']V^*(S')\bigg].
	\end{aligned}
	\label{eq:scomdp}
\end{equation}
After that, $\theta^* = \min\limits_{\Omega}\theta(\Omega)$ is the optimal average cost for any initial state $S(0)$. Suppose there exists a stationary admissible $\Omega^*$ with $\Omega^*(S) = \{P^*_l, P^*_t\}$ for any $S$, where $\{P^*_l, P^*_t\}$ attains the minimum of the R.H.S. in (\ref{eq:scomdp}) for given $S$. The optimal offloading policy of Problem \ref{prob:dopco} is achieved by $\Omega^*$.
Then, taking the expectation w.r.t. $H$ on both sizes of (\ref{eq:scomdp}), and denoting $V^*(Q_l, Q_r) = \mathbb{E}[V^*(S)|Q_l, Q_r]$. Finally we obtain the equivalent Bellman equation in (\ref{eq:bellman}) in Theorem \ref{th:SCO}.

\section*{Appendix B: Proof of Theorem \ref{th:HJB}}
Suppose $V(q_l, q_r)$ is of class $\mathcal{C}^1(\mathbb{R}_+^2)$, we have $dV(q_l, q_r) = \frac{\partial V(q_l, q_r)}{\partial q_l}dq_l + \frac{\partial V(q_l, q_r)}{\partial q_r}dq_r$. Substituting the dynamics in (\ref{eq:dq}), we obtain
\begin{equation}
\begin{aligned}
	dV(q_l(t), &q_r(t)) = \Delta ^{\Omega ^v}\big(V(q_l(t), q_r(t))\big)dt + \frac{\partial V(q_l(t), q_r(t))}{\partial q_l}dRe_l(t) \\
	&+ \bigg[\frac{\partial V(q_l(t), q_r(t))}{\partial q_l}-\frac{\partial V(q_l(t), q_r(t))}{\partial q_r}\bigg]dRe_t(t) + \frac{\partial V(q_l(t), q_r(t))}{\partial q_r}dRe_r(t)
\end{aligned}
\label{eq:dv}
\end{equation}
where $\Delta ^{\Omega ^v}\big(V(q_l, q_r)\big) = \frac{\partial V(q_l, q_r)}{\partial q_l}\bigg[\overline{\lambda} - v_l(P_l) - \mathbb{E}\big[v_t(P_t, H)\big| q_l, q_r\big]\bigg]
	+ \frac{\partial V(q_l, q_r)}{\partial q_r}\bigg[\mathbb{E}\big[v_t(P_t, H)\big| q_l, q_r\big] - \overline{v_{out}}\bigg]$.
Integrating (\ref{eq:dv}) on both sizes w.r.t. $t$ from $0$ to $T$, we obtain
\begin{align}
	V(q_l(T), &q_r(T)) - V(q_l(0), q_r(0)) \nonumber\\
=&\int_{0}^{T}\bigg[\Delta ^{\Omega ^v}(V(q_l(t), q_r(t))) + \int_{0}^{T}\frac{\alpha}{\overline{\lambda}}\big(q_l(t) + q_r(t)\big) + \beta \big(P_l(t) + P_t(t)\big) \bigg]dt\nonumber\\
	&+ \int_{0}^{T}\frac{\partial V(0, q_r(t))}{\partial q_l}dRe_l(t) + \int_{0}^{T}\bigg[\frac{\partial V(0, q_r(t))}{\partial q_l} - \frac{\partial V(0, q_r(t))}{\partial q_r}\bigg]dRe_t(t)\nonumber\\
	&+ \int_{0}^{T}\frac{\partial V(q_l(t), 0)}{\partial q_r}dRe_r(t) - \int_{0}^{T}\bigg[\frac{\alpha}{\overline{\lambda}}\big(q_l(t) + q_r(t)\big) + \beta \big(P_l(t) + P_t(t)\big) \bigg]dt
\label{eq:v_int1}
\end{align}
where $Re_l(t)$ and $Re_t(t)$ increase only when $q_l = 0$, and $Re_r(t)$ increases only when $q_r = 0$ according to the definition of reflection. If $V(q_l, q_r)$ satisfies (\ref{eq:orihjb}), from (\ref{eq:v_int1}), for any admissible virtual policy $\Omega ^v$, we have
\begin{align}
	V(q_l(0), q_r(0)) \leq & V(q_l(T), q_r(T)) - \int_{0}^{T}\frac{\partial V(0, q_r(t))}{\partial q_l}dRe_l(t) \nonumber\\
	& - \int_{0}^{T}\bigg[\frac{\partial V(0, q_r(t))}{\partial q_l}- \frac{\partial V(0, q_r(t))}{\partial q_r}\bigg]dRe_t(t) + \int_{0}^{T}\frac{\partial V(q_l(t), 0)}{\partial q_r}dRe_r(t)\nonumber\\
	& +\int_{0}^{T}\bigg[\frac{\alpha}{\overline{\lambda}}\big(q_l(t) + q_r(t)\big) + \beta \big(P_l(t) + P_t(t)\big)\bigg]dt- C^{\infty}T
\label{eq:v_int2}
\end{align}
From the boundary conditions in (\ref{eq:bound_cond}), taking the limit superior as $T\to \infty$ in (\ref{eq:v_int2}), we have
\begin{equation}
	\frac{1}{T}V(q_l(0), q_r(0)) \leq \frac{1}{T}\limsup_{T\to \infty} \int_{0}^{T}\bigg[\frac{\alpha}{\overline{\lambda}}\big(q_l(t) + q_r(t)\big) + \beta \big(P_l(t) + P_t(t)\big)\bigg]dt - C^{\infty}.
\end{equation}
Because $V(q_l(0), q_r(0))$ is bounded, We have
\begin{equation}
C^{\infty} \leq \frac{1}{T}\limsup_{T\to \infty} \int_{0}^{T}\bigg[\frac{\alpha}{\overline{\lambda}}\big(q_l(t) + q_r(t)\big) + \beta \big(P_l(t) + P_t(t)\big)\bigg]dt = \theta(q_l(0), q_r(0);\Omega ^ v),
\end{equation}
where the above equality is achieved if the admissible virtual stationary offloading policy $\Omega ^v(H, q_l, q_r)$ attains the minimum in the HJB equation in (\ref{eq:orihjb}) for all $(H, q_l, q_r)$. Hence, such $\Omega ^v$ is the optimal offloading policy of the average cost problem in VCTS in Problem \ref{prob:vcts}.

\section*{Appendix C: Proof of Theorem \ref{th:simhjb}}
First, we simplify the PDE in (\ref{eq:orihjb}). Based on Theorem~\ref{coro:ptstar}, the optimal offloading policy that minimizes the L.H.S of (\ref{eq:orihjb}) is given by (\ref{eq:plstar}) and (\ref{eq:ptstar}). Substituting them to the PDE in (\ref{eq:orihjb}), and denoting $\mathrm{EP}(x)\triangleq \mathbb{E}[P^*_t]=\int_{\frac{\beta N_0}{xB}}^{+\infty}\frac{1}{L}e^{-\frac{H}{L}}\big(\frac{xB}{\beta}-\frac{N_0}{H}\big)dH=\frac{xB}{\beta}\exp\big(-\frac{\beta N_0}{xBL}\big) - \frac{N_0}{L}E_1\big(\frac{\beta N_0}{xBL}\big)$ and $\mathrm{EVP}(x) \triangleq \mathbb{E}[v_t(P^*_t, H)]=\int_{\frac{\beta N_0}{xB}}^{+\infty}\frac{1}{L}e^{-\frac{H}{L}}B\log_2\big(\frac{xBH}{\beta N_0}\big)dH=BE_1\big(\frac{\beta N_0}{xBL}\big)$ according to the Rayleigh distribution of $H$, the PDE in (\ref{eq:orihjb}) can be simplified into (\ref{eq:simhjb}).

%

For the continuous time queue system in (\ref{eq:dq}), there exist the steady data queue states $q_{l,s}\triangleq \lim\limits_{t\to \infty}q_l(t) =0$ and $q_{r,s} \triangleq \lim\limits_{t\to \infty}q_r(t)=0$. 
At the steady states, the average departure rate should be larger than or equal to the average arrival rate, i.e.,
\begin{equation}
\begin{aligned}
	&\overline{\lambda} \leq \mathrm{EVP}(x) + \frac{\overline{k}^2}{2c\beta}V_l&\\
	&\mathrm{EVP}(x) \leq \overline{v_{out}}&
\end{aligned}
\label{eq:raterange}
\end{equation}

In the following lemma, we exclude the case with $\overline{\lambda} < \mathrm{EVP}(x) + \frac{\overline{k}^2}{2c\beta}V_l$ from the steady states.
\begin{lemma}[Feasibility of Steady State]	
	 There does not exist a feasible solution of $V(q_l, q_r)$ that satisfies (\ref{eq:orihjb}) and (\ref{eq:bound_cond}) if $\overline{\lambda} < \mathrm{EVP}(x) + \frac{\overline{k}^2}{2c\beta}V_l$.
\end{lemma}
\begin{IEEEproof}
	At the steady state, the queue states satisfy $q_{l,s} = 0$ and $q_{r,s} = 0$. If $\overline{\lambda} < \frac{\overline{k}^2}{2c\beta}V_l + \mathrm{EVP}(x)$, based on the definition of reflection, $dRe_l(t) \neq 0$ and $dRe_t(t)\neq 0$.
	Considering the boundary condition (\ref{eq:bound_cond}), the solution of $V(q_l, q_r)$ should satisfy
	\begin{equation}
		\frac{\partial V(0, q_r)}{\partial q_l} = 0,
	\end{equation}
	\begin{equation}
		x(0, q_r) = \frac{\partial V(0, q_r)}{\partial q_l} - \frac{\partial V(0, q_r)}{\partial q_r} = 0.
	\end{equation}
	Thus, with the offloading policy at the steady state,
		$\mathrm{EVP}(x) + \frac{\overline{k}^2}{2c\beta}V_l = 0 < \overline{\lambda}$,
	which leads to a contrary.
\end{IEEEproof}
Therefore, the sufficient conditions of the existence of solution in Theorem~\ref{th:simhjb} is obtained.

\section*{Appendix D: Proof of Theorem \ref{th:SS}}
{To achieve the optimal offloading policy, the steady state should satisfy two criteria: queue stability and average cost optimality. Specifically, we can obtain the optimal steady state by solving the following convex optimization problem, which aims to optimize the average cost with the queue stability constraints:}
\begin{eqnarray}
	&\min_{x, V_l} &G(x, V_l) = \beta \mathrm{EP}(x) + \frac{\overline{k}^2}{4c\beta}V^2_l
	\label{eq:apd1}\\
	&\mathrm{s.t.}&\overline{\lambda}=\mathrm{EVP}(x)+\frac{\overline{k}^2}{2c\beta}V_l
	\label{eq:apd2}\\
	&&\mathrm{EVP}(x) \leq \overline{v_{out}}\\
	&&0 \leq x \leq V_l
\end{eqnarray}

Substituting (\ref{eq:apd2}) into (\ref{eq:apd1}), we obtain
$G(x) = \beta \mathrm{EP}(x) + \frac{c\beta}{\overline{k}^2}\big[\overline{\lambda} - BE_1\big(\frac{\beta N_0}{xBL}\big)\big]^2$,
Since $\frac{dG}{dx}=-\frac{N_0}{xBL}\exp(-\frac{\beta N_0}{xBL}) - \frac{2c\beta}{x\overline{k}^2}\exp(-\frac{\beta N_0}{xBL})\big[\overline{\lambda} - BE_1\big(\frac{\beta N_0}{xBL}\big)\big] \leq 0$,
$G$ is a non-increasing function of $x$. Considering that $x_e$ satisfies $\mathrm{EVP}(x_e) = \overline{v_{out}}$, we have $x = \min\{V_l, x_e\}$.

If $\overline{\lambda} < \mathrm{EVP}(x_e) + \frac{\overline{k}^2}{2c\beta}x_e$, the optimal solution $x < x_e$, so $x = V_l$ and the steady state should satisfy $\overline{\lambda} = \mathrm{EVP}(V_l) + \frac{\overline{k}^2}{2c\beta}V_l,\;\mathrm{EVP}(V_l) < \overline{v_{out}}$. Else, the performance is limited by the computation ability of the MEC server, so the optimal solution is $x = x_e$ and the steady state should satisfy $\overline{\lambda} = \mathrm{EVP}(x) + \frac{\overline{k}^2}{2c\beta}V_l,\;\mathrm{EVP}(x) = \overline{v_{out}}$.

\section*{Appendix E: Proof of Theorem \ref{th:closedform1}}

We first rewrite (\ref{eq:simhjb}) as
\begin{equation}
\frac{\alpha}{\overline{\lambda}}\big(q_l + q_r\big) + \beta \mathrm{EP}(x) - x\mathrm{EVP}(x) + V_l\overline{\lambda} - \frac{\overline{k}^2}{4c\beta}V^2_l - V_r \overline{v_{out}} - C^{\infty} = 0
\label{eq:finalhjb}
\end{equation}

Based on Theorem \ref{def:del}, we have the following approximations in (\ref{eq:finalhjb}):
\begin{align}
	&\beta\mathrm{EP}(x) = B\exp\bigg(-\frac{\beta N_0}{V_{l,c}BL}\bigg)V_l - \frac{\beta N_0}{L}E_1\bigg(\frac{\beta N_0}{V_{l,c}BL}\bigg) + o(V_l - V_{l,c})^2,\label{eq:apprep}\\
	&x\mathrm{EVP}(x) = \bigg[B\exp\bigg(-\frac{\beta N_0}{V_{l,c}BL}\bigg) + BE_1\bigg(\frac{\beta N_0}{V_{l,c}BL}\bigg)\bigg]V_l \nonumber\\
	&\qquad\qquad\qquad\qquad\qquad\qquad- V_{l,c}B\exp\bigg(\frac{\beta N_0}{V_{l,c}BL}\bigg) + o(V_l - V_{l,c})^2,\label{eq:apprevp}\\
	&\frac{\overline{k}^2}{4c\beta}V^2_l = \frac{\overline{k}^2}{2c\beta}V_{l,c}V_l - \frac{\overline{k}^2}{4c\beta}V^2_{l,c} + o(V_l-V_{l,c})^2.
	\label{eq:apprvl}
\end{align}
Substituting (\ref{eq:apprep}), (\ref{eq:apprevp}), and (\ref{eq:apprvl}) into (\ref{eq:finalhjb}), we obtain the following simplified PDE:
\begin{equation}
\epsilon V_l + \bigg[\overline{v_{out}} - BE_1\big(\frac{\beta N_0}{V_{l,c}BL}\big)\bigg]V_r = \frac{\alpha}{\overline{\lambda}}\big(q_l + q_r\big) + C - C^{\infty}.
\end{equation}
Using \emph{3.2.1.2} of \cite{AD2002}, we obtain the solution of the above PDE as
\begin{equation}
	V(q_l, q_r) = \frac{\alpha}{2\overline{\lambda}\epsilon}q^2_l + \frac{C-C^{\infty}}{\epsilon}q_l + \frac{\alpha}{2\overline{\lambda}\big[\overline{v_{out}} - BE_1\big(\frac{\beta N_0}{V_{l,c}B L}\big)\big]}q^2_r + \Phi\bigg(\bigg[\overline{v_{out}} - BE_1\big(\frac{\beta N_0}{V_{l,c}B L}\big)\bigg]q_l - \epsilon q_r\bigg).
\end{equation}

Next, we determine the function $\Phi(*)$ and other undetermined constants in the solution according to the boundary conditions in (\ref{eq:bound_cond}).
To satisfy the last condition of (\ref{eq:bound_cond}), we choose $\Phi(*) = 0$. 
Based on the expression of $V_r$, the third condition of (\ref{eq:bound_cond}) is satisfied.
Under the steady state, $\overline{\lambda} = \mathrm{EVP}(V_l) + \frac{\overline{k}^2}{2c\beta}V_l$ according to Theorem~\ref{th:SS}.
Let $q_l = q_r = 0$, then we have $V_l = \frac{C - C^{\infty}}{\epsilon}$. Because $\epsilon > 0$, $V_l \in (V_{l,s}, V_{l,c})$. We have
\begin{equation}
	\epsilon(t+1) = f(\frac{C(t)-C^{\infty}}{\epsilon(t)}) < f(V_{l,c}) = \epsilon(t)
\end{equation}
where $f(V) = \mathrm{EVP}(V)+\frac{\overline{k}^2}{2c\beta}V - \overline{\lambda}$. 
Thus, $\epsilon$ will be convergent to 0 and the steady state is $V_{l,c} = V_{l,s}$. Therefore, the first two conditions in (\ref{eq:bound_cond}) are satisfied.



\section*{Appendix F: Proof of Theorem \ref{th:apprscincsc}}
%
%
%
%
%

	Based on Definition \ref{def:del} and Theorem \ref{th:closedform1}, when $\epsilon$ tends to 0, we have
	\begin{equation}
		O(\epsilon) = O(\ln(V_{l,c})) + O(V_{l,c}) = O(V_{l,c}),
	\end{equation}
	\begin{equation}
		O(C - C^{\infty}) = O(\ln(V_{l,c})) + O(V^2_{l,c}) = O(\epsilon^2),
	\end{equation}
	so the error between the steady state is
	\begin{equation}
		\frac{C - C^{\infty}}{\epsilon} - V_{l,s} = \frac{O(\epsilon^2)}{O(\epsilon)} = O(\epsilon).
	\end{equation}
	When $\epsilon = \epsilon_0$, the approximation error is $O(\epsilon_0)$.


\section*{Appendix G: Proof of Theorem \ref{th:closedform2}}
Similar to Theorem \ref{th:closedform1}, we can obtain
\begin{equation}
V(q_l, q_r) = \frac{\alpha}{2\overline{\lambda}\epsilon}q^2_l + \gamma\frac{C-C^{\infty}}{\epsilon}q_l + (1-\gamma)\frac{C-C^{\infty}}{\delta}q_r + \frac{\alpha}{2\overline{\lambda}\delta}q^2_r + \Phi(\delta q_l - \epsilon q_r).
\end{equation}
To satisfied the last condition of (\ref{eq:bound_cond}), we choose $\Phi(*) = 0$. Then we obtain the expressions of $V_l$ and $V_r$.
\begin{eqnarray}
V_l &=& \frac{\alpha}{\overline{\lambda} \epsilon}q_l + \gamma\frac{C-C^{\infty}}{\epsilon}\\
V_r &=& \frac{\alpha}{\overline{\lambda} \delta}q_r + (1 - \gamma)\frac{C-C^{\infty}}{\delta}
\end{eqnarray}
We let $q_l = q_r = 0$, then $V_l = \gamma \frac{C - C^{\infty}}{\epsilon}$, $V_r = (1 - \gamma)\frac{C - C^{\infty}}{\delta}$. Next, we need to determine the value of $\gamma$ to guarantee the system can be convergent to the steady state. We obtain the $\gamma$ from the following convex optimization problem.
\begin{align}
\max~~&\bigg[x_s - \big[\gamma \frac{C - C^{\infty}}{\epsilon} - (1 - \gamma)\frac{C - C^{\infty}}{\delta}\big]\bigg]\bigg[\gamma \frac{C - C^{\infty}}{\epsilon}-V_{l,s}\bigg]\\
\mathrm{s.t.}~~& \gamma \in \Gamma
\end{align}

The extreme point of $\gamma$ is $\gamma = \frac{(x_s + V_{l,s})\epsilon\delta}{2(\epsilon + \delta)(C-C^{\infty})} + \frac{V_{l,s}\epsilon^2}{2(\epsilon + \delta)(C-C^{\infty})} + \frac{\epsilon}{2(\epsilon + \delta)}$, which can obtain the maximum balance of convergence. Based on different value of $\epsilon$ and $\delta$, the constraint will affect the feasible region. We choose the nearest value of extreme point for $\gamma$. Like the proof of Theorem \ref{th:closedform1}, we can proof the steady state is achieved, and the proof is omitted for brevity.

\section*{Appendix H: Proof of Theorem \ref{th:stable}}
First, we try to prove that under a sufficiently large queue $Q_l(0)$, the local queue has a negative drift.
One step drift of the local queue is calculated as
\begin{equation}
\begin{aligned}
	&\mathbb{E}\bigg[Q_l(n+1) - Q_l(n)\bigg|Q_l(n) = Q'_l,\; Q_r(n) = Q'_r\bigg]\\
	= &\mathbb{E}\bigg[\big[Q'_l - v_t(P^*_t, H)\tau - v_l(P^*_l)\tau\big]^+ + \lambda \tau - Q'_l\bigg]\\
	\overset{(a)}{=}& \mathbb{E}\bigg[- v_t(P^*_t, H)\tau - v_l(P^*_l)\tau + \lambda \tau\bigg]\\
	=& \overline{\lambda}\tau - \mathbb{E}\bigg[v_t(P^*_t, H) + v_l(P^*_l)\bigg]\tau \overset{(b)}{<}0,
\end{aligned}
\end{equation}
where $(a)$ is due to for a large local queue and a small slot $\tau$, we have $\mathrm{Pr}[Q'_l > v_t(P^*_t, H)\tau + v_l(P^*_l)\tau] > {1 - \xi\;(\forall \xi > 0)}$. $(b)$ is due to the output rate of local queue is larger than the input rate when $Q_l(n) > 0$. Hence, the local queue has a negative drift. 

Next, we try to prove the negative drift for the remote queue.
In computation sufficient scenario, the negative drift is obvious and we omit the proof for brevity. We mainly consider the computation constrained scenario. Similar to the proof for the local queue, we have
\begin{equation}
\begin{aligned}
	&\mathbb{E}\bigg[Q_r(n+1) - Q_r(n)\bigg|Q_l(n) = Q'_l,\; Q_r(n) = Q'_r\bigg]\\
	=&\mathbb{E}\bigg[\big[Q'_r - v_{out}\tau\big]^+ + v_t(P^*_t, H)\tau - Q'_r\bigg]\\
	\overset{(c)}{=}& \mathbb{E}\bigg[-v_{out}\tau + v_t(P^*_t, H)\tau\bigg]\\
	=& \mathbb{E}\bigg[v_t(P^*_t, H)\bigg]\tau - \overline{v_{out}}\tau \overset{(d)}{<}0,
\end{aligned}
\end{equation}
where $(c)$ and $(d)$ is similar to $(a)$ and $(b)$. Therefore, we prove that the remote queue has a negative drift.




%


\end{document}